\documentclass[12pt]{article}
\usepackage{amsmath, amsfonts, graphicx, graphics, bm, float, adjustbox, lscape, threeparttable, rotating, makecell, setspace, tabularx, multirow, booktabs, natbib, caption, subcaption, chngcntr, dsfont}
\usepackage[table,xcdraw]{xcolor}
\usepackage{tikz}
\usetikzlibrary{positioning}
\usepackage[utf8]{inputenc}
\usepackage[top=3cm,left=3cm,right=3cm,bottom=3cm]{geometry}
\usepackage[graphicx]{realboxes}
\usepackage [english]{babel}
\usepackage [autostyle, english = american]{csquotes}
\usepackage[all,cmtip]{xy} 
\MakeOuterQuote{"}
\usepackage[toc,page]{appendix}
\usepackage[T1]{fontenc}
\usepackage{hyperref}
\usepackage{authblk}
\hypersetup{
colorlinks=true,
citecolor=blue,
linkcolor=blue,
filecolor=blue,      
urlcolor=blue,
}

\usepackage{ulem}
\usepackage{array}
\newcolumntype{C}[1]{>{\centering\arraybackslash}p{#1}}
\newcolumntype{L}[1]{>{\raggedright\arraybackslash}m{#1}}
\newcolumntype{M}[1]{>{\centering\arraybackslash}m{#1}}

\graphicspath{{pictures}}

\begin{document}

\begin{titlepage}
\title{\Large Owning the Intelligence: Global AI Patents Landscape and Europe’s Quest for Technological Sovereignty}

\author[1,2]{Lapo Santarlasci\thanks{Corresponding author. Email: lapo.santarlasci@imtlucca.it. CRediT: Conceptualization, Methodology, Software, Data curation, Formal analysis, Validation, Investigation, Visualization, Resources, Writing – original draft.}}
\author[1]{Armando Rungi\thanks{CRediT: Conceptualization, Investigation, Methodology, Supervision, Writing – review \& editing, Resources.}}
\author[2]{Loredana Fattorini\thanks{CRediT: Conceptualization, Investigation, Validation, Methodology, Supervision, Writing – review \& editing, Resources, Project administration.}}
\author[2]{Nestor Maslej\thanks{CRediT: Conceptualization, Validation, Supervision, Resources, Project administration.}}

\affil[1]{\footnotesize Laboratory for the Analysis of Complex Economic Systems (AXES), IMT School for Advanced Studies Lucca, Lucca, Italy}
\affil[2]{\footnotesize Institute for Human-Centered AI, Stanford University, Stanford, United States}

\date{}
\maketitle
\vspace{0.1in}
\begin{abstract}
\footnotesize
Artificial intelligence has become a key arena of global technological competition and a central concern for Europe’s quest for technological sovereignty. This paper analyzes global AI patenting from 2010 to 2023 to assess Europe’s position in an increasingly bipolar innovation landscape dominated by the United States and China. Using linked patent, firm, ownership, and citation data, we examine the geography, specialization, and international diffusion of AI innovation.

We find a highly concentrated patent landscape: China leads in patent volumes, while the United States dominates in citation impact and technological influence. Europe accounts for a limited share of AI patents but exhibits signals of relatively high patent quality. Technological proximity reveals global convergence toward U.S. innovation trajectories, with Europe remaining fragmented rather than forming an autonomous pole. Gravity-model estimates show that cross-border AI knowledge flows are driven primarily by technological capability and specialization, while geographic and institutional factors play a secondary role. EU membership does not significantly enhance intra-European knowledge diffusion, suggesting that technological capacity, rather than political integration, underpins participation in global AI innovation networks.\\

\vspace{0cm}
\bigskip
\end{abstract}
\setcounter{page}{0}
\thispagestyle{empty}
\end{titlepage}
\pagebreak \newpage

\onehalfspacing

\maketitle

\section{Introduction}
\label{sec: intro}

Artificial intelligence technologies have advanced rapidly in recent years, attracting growing attention from scholars, policymakers, and industry leaders \citep{agrawal_economics_2019, giczy_identifying_2021}. Recognized as a general-purpose technology - akin to the steam engine, electricity, or the internet - AI is expected to reshape production and innovation across diverse domains \citep{brynjolfsson_what_2018, wipo_wipo_2019, dernis_world_2019, igna_determinants_2023}. Patent data confirm this wide-ranging impact, showing AI applications well beyond the IT and electronics sectors, extending into telecommunications, transport, healthcare (e.g., disease diagnosis and drug discovery), finance, manufacturing, and agriculture \citep{dernis_world_2019, wipo_wipo_2019, desouza_designing_2020, giczy_identifying_2021}.
AI enhances innovation by augmenting or automating tasks such as decision-making, design, and abductive reasoning \citep{gama_artificial_2025}. It enables firms to overcome information constraints, mine large R\&D datasets, and identify promising technological trajectories \citep{haefner_artificial_2021, mariani_artificial_2023}. These capabilities position AI as a transformative force in how firms innovate and compete, making research on patent production in this technological domain and the profiles of their producers increasingly important for understanding the evolving landscape of technological change. \\

\noindent This paper provides a comprehensive analysis of global AI patenting activity between 2010 and 2023, with particular attention to Europe’s position within emerging international innovation dynamics, and a focus on the organizational and international dimensions of innovation. The rhetoric of sovereignty has infact increasingly become central to the evolving framework of European (EU) industrial policy, functioning both as a legitimizing discourse and as a structuring principle for new coalitional alignments among member states \citep{seidl_moving_2024}. In this context, Europe’s relative dependence on external actors in several critical technologies, including artificial intelligence, has revived debates on technological sovereignty. As emphasized by \cite{crespi_european_2021}, the relaunch of a European industrial policy should explicitly embed technological sovereignty as a guiding objective-defining priorities, setting strategic targets, and shaping the design of policy instruments accordingly. From this perspective, technological sovereignty serves not merely as a defensive notion of autonomy, but as an overarching policy framework that integrates innovation capacity, industrial competitiveness, and strategic resilience. Understanding Europe’s positioning in the global AI patent landscape therefore contributes to assessing how far the continent has advanced toward such a goal, and to what extent EU policy initiatives can strengthen its technological self-determination within an increasingly bipolar global system. Beyond its policy relevance, this perspective frames our empirical inquiry: to what extent does Europe exhibit distinctive patterns of specialization, collaboration, and knowledge diffusion in AI patenting compared with the global technological leaders, notably the United States and China? Addressing this question allows us to assess whether Europe’s AI innovation system operates as an autonomous pole, a transatlantic partner, or a peripheral follower in the emerging bipolar configuration of global AI knowledge production. \\

\noindent We build on the dataset of AI-related patents compiled by the Stanford AI Index team \citep{maslej_artificial_2025}, and enrich this data by linking it to firm-level and ownership information from ORBIS Intellectual Property, thereby assigning AI patents to their applicants and parent company. This integration allows us to move beyond aggregate counts and provide a granular picture of the geography, sectors, and organizations driving AI patented innovation worldwide. A further contribution of this paper is to examine how AI-related knowledge diffuses across countries by estimating a gravity model of bilateral patent citations. This approach, drawing on the tradition of gravity-based analyses of trade and technological spillovers, allows us to quantify how economic size, technological specialization, geographic frictions, and institutional proximity shape the international transmission of AI inventions. By linking countries’ AI patenting intensity to their citation exchanges, we identify the structural drivers of cross-border knowledge flows and assess whether Europe behaves as an integrated innovation area or as a collection of heterogeneous national systems. \\

\noindent Our findings highlight six main dimensions of the contemporary AI patent landscape. First, AI patented innovation is highly concentrated: China is the leading country in terms of patent counts, while the United States dominates in citation impact, underscoring its technological influence. Europe play a marginal role in terms of granted patents and citations, with an aggregated portfolio smaller even than the one of Japan. Second, the dual role of national firms and multinational corporations is evident: while the majority of applicants are domestic firms, multinational companies-though fewer in number-account for the majority of granted patents. Third, sectoral patterns reveal the centrality of ICT, manufacturing, and education, with universities and research institutions contributing disproportionately to AI patented innovation. Fourth, our measured technological proximity highlight patented innovation patterns globally closer to the U.S. rather than China, with EU countries notably scattered around different trajectories. Fifth, our measures of revealed comparative advantage (RCA) provide a nuanced view of national specialization, showing that Europe, as well as absolute leaders like China and Japan, is not relatively specialized in AI, while smaller single countries such as Israel, and Ireland display a strong AI focus. Finally, our results provide novel evidence on the determinants of AI knowledge diffusion across countries. Using a gravity model framework adapted to bilateral AI patent citations, we find that while geographic and cultural proximity still shape international knowledge flows, their relative importance declines once countries’ AI innovation capacity and technological proximity are introduced. The elasticity of AI patenting activity (around 0.7\%) suggests that technologically advanced and specialized systems are substantially more likely to exchange knowledge across borders. However, belonging to the European Union does not exert a significant independent effect on citation intensity once these technological factors are accounted for. This finding implies that Europe’s regional integration has yet to translate into a cohesive innovation system for AI. Taken together, the results depict an increasingly capability-driven global landscape, where Europe remains an intermediate player - competitive in technological quality, yet still fragmented and dependent on external poles of innovation. These insights not only inform our understanding of AI’s international diffusion but also speak directly to current European policy debates on industrial and technological sovereignty. \\

\noindent Our results contribute to the literature in two ways. Empirically, we provide the most detailed firm-level mapping of AI patents to date, distinguishing the role of organizational types, national versus multinational ownership, and sectoral activity. Analytically, we introduce new perspectives by applying RCA, technological proximity, and a gravity model of knowledge spillovers to the domain of AI, thereby uncovering structural patterns in specialization and diffusion. At the same time, our descriptive evidence on geographic concentration, sectoral dominance, citation leadership, and home biases is consistent with prior studies, reinforcing established stylized facts while extending them to the AI domain. \\

\noindent The remainder of the paper is structured as follows. Section \ref{sec: literature} reviews the existing literature on the identification of AI patents and the determinants of AI innovation. Section \ref{sec: data_and_method} describes the data sources and methodology. Section \ref{sec: descriptive_evidence} presents descriptive evidence on the geography, sectoral distribution, leading players in AI patenting, and technological proximity. Section \ref{sec: rca} evaluates countries’ revealed comparative advantage in AI. Section \ref{sec: impact} analyzes the impact and international spillovers of AI patents, including citation patterns, and gravity model estimates. Section \ref{sec: conclusion} concludes with a discussion of the broader implications for innovation policy and global competition in the AI era.

\section{Related literature}
\label{sec: literature}

Our paper contributes to two different strands of literature. First it primary contributes to mapping and analytically describing the landscape of AI patents global production and original innovators, e.g., \cite{tseng_patent_2013, fujii_trends_2018, agrawal_economics_2019, alderucci_quantifying_2019, benassi_rush_2020, cockburn_impact_2018, damioli_impact_2021, igna_determinants_2023, mariani_artificial_2023, parteka_artificial_2023}, an increasing studied area of a recently developed strand of economics of innovation centered on AI, e.g., \cite{verganti_innovation_2020, haefner_artificial_2021, venturini_intelligent_2022, halaburda_business_2024, gama_artificial_2025}. Second we can comment on regional specialization, e.g., \cite{arenal_innovation_2020}, and knowledge spillover aspects of the AI patented innovation landscape, e.g., \cite{jaffe_geographic_1993, jaffe_technological_1986, bar_measure_2012}. \\

\noindent Early efforts to identify AI patents primarily relied on patent classification systems - e.g. \cite{tseng_patent_2013, cockburn_impact_2018} and \cite{fujii_trends_2018}. These methods are systematic and reproducible, but suffer from limitations: classification schemes lag behind emerging technologies and may be inconsistently applied, missing novel AI applications. To address these gaps, researchers adopted keyword-based text mining - e.g. \cite{van_roy_ai_2019} and \citep{wipo_wipo_2019}. These approaches allow timely updates and better detection of emerging terms but risk high false positive rates due to ambiguity, language variation, and the evolving nature of AI terminology. The limitations of both classification and keyword approaches prompted a turn to machine learning, e.g.: \cite{alderucci_quantifying_2019, giczy_identifying_2021, miric_using_2023, pairolero_artificial_2024} and  \cite{wipo_patent_2024}. These techniques boost accuracy and semantic depth but require extensive resources and training data, and often suffer from interpretability issues. More recently, hybrid approaches have gained traction by combining the strengths of previous methods. \cite{benassi_rush_2020, igna_determinants_2023} and \cite{maslej_artificial_2025}. \\

\noindent Accurate identification of AI patents is not only a methodological concern but a prerequisite for empirical analysis of AI's broader economic and organizational impact. As patent data serve as a proxy for technological capabilities and innovation strategies, refined methods for detecting AI-related inventions enable researchers to explore how AI adoption correlates with firm-level outcomes such as productivity, employment, and competitiveness. This has led to a growing body of work leveraging AI patent datasets to investigate how AI reshapes firm behavior, industry dynamics, and national innovation systems. 
\cite{alderucci_quantifying_2019} showed that AI-patenting firms in the U.S. are larger, older, and more productive, with higher employment and income inequality. \cite{damioli_impact_2021} found AI adoption in Europe improved labor productivity, especially for SMEs and service firms. \cite{van_roy_ai_2019} examined how AI patenting varies across sectors and its correlation with firm performance. Collectively, these studies suggest that AI innovation is both a marker and a driver of firm-level competitiveness. \\

\noindent Organizational and policy determinants of AI patenting have also been studied. \cite{igna_determinants_2023} linked AI innovation to human capital, collaboration, and technological capabilities. \cite{fujii_trends_2018} emphasized the role of R\&D policies and industrial characteristics in shaping AI invention, showing how firm strategies and national contexts influence both the direction and scale of AI activity. Furthermore, some studies found out that, spatially, AI innovation is highly concentrated. \cite{miric_using_2023} identified U.S. hotspots like California, New York, and Washington. \cite{de_ai_2019} compared institutional players across countries, while \cite{tseng_patent_2013} analyzed national strengths and cross-border ownership in AI subfields. Research has also distinguished between organizational types. \cite{cockburn_impact_2018} tracked the rising role of private firms post-2009, while \cite{benassi_rush_2020} used clustering models to identify specialization patterns, and \cite{righi_ai_2022} highlighted how firms and research institutes play complementary roles in AI innovation. \\

\noindent Finally, competitive dynamics have been explored through the analysis of patent portfolios. \cite{wipo_wipo_2019} highlighted dominant tech firms and the rise of Chinese actors. \cite{murdick_patents_2020} introduced citation-based quality measures and identified new entrants among public and academic institutions. \cite{dernis_world_2019} focused on top R\&D investors, showing sectoral differences in AI specialization. Recently, \cite{wipo_patent_2024} presented the first in-depth patent analysis of Generative AI, capturing trends across models and modalities. \\

\noindent In sum, the existing literature has made important progress both in refining the identification of AI-related patents and in exploring their organizational, regional, and economic determinants. Yet most contributions remain either methodologically focused or limited to aggregate patterns, leaving less attention to firm-level structures, ownership linkages, and the interaction between absolute leadership and relative specialization. Moreover, while some studies address geographic clustering and citation dynamics, systematic analyses of international spillovers in AI remain scarce. Our paper seeks to bridge these gaps by combining a robust hybrid dataset of AI patents with detailed firm- and ownership-level information, enabling us to uncover both the organizational forms and international dimensions of AI patented innovation.

\section{Data and methodology}
\label{sec: data_and_method}
Our methodology combines several data sources to create a comprehensive firm-level dataset of AI\footnote{Robotics and automation technologies are also considered only when AI-based.} patents filed between the year 2010 and 2023. We rely on data Stanford annual AI Index Report \citep{maslej_artificial_2025}. The AI Index team utilizes bibliographic patent data from the European Patent Office's Worldwide Patent Statistical Database (PATSTAT). To construct their dataset of AI-related patents, the authors implemented a comprehensive, two-pronged classification methodology, which integrated both a keyword-based text analysis and a classification-code-based approach, further validated by rigorous mechanisms. The dataset contains, for each granted AI patent (and related publications): patent authority, classification codes (IPCs and CPCs), granted date, earliest publication date and forward citations. 
We merged\footnote{Please note that 24 granted applications are not present in ORBIS IP database; however those account for less than 0.006\% of total AI patents during the sample period.} this information with firm-level data\footnote{Please note that for around 17\% of AI patents during the period considered there is no available information on the applicant in ORBIS IP.} from the ORBIS Intellectual Property database, compiled by Bureau van Dijk (please refer to Appendix \ref{sec:appendix_tab_graph}, Table \ref{tab: missing_orbis} for information on firm-level missing data). This crucial step allowed us to link patents to their specific applicants - firms, universities, and other entities driving AI patented innovation. The ORBIS IP data provides detailed information on each applicant, including their country of incorporation, primary industry classification (NACE rev.2), and year of establishment, together with information on financial accounts. \\ 

\noindent In addition, we leverage the ownership information within ORBIS to identify the parent company for each applicant. This allows us to distinguish between the geographical location of the patenting subsidiary and the headquarters of the parent company, providing a more nuanced view of where control over patent activity resides. The final dataset comprises a comprehensive list of AI patents granted between 2010 and 2023, linked to the characteristics of the applying firms and their ultimate owners.
Finally, to analyze Europe’s position in the global landscape, we complement the country-level indicators with EU-level aggregates: for all EU member states, we construct a combined AI patent portfolio that enables us to treat Europe as a single innovation area when examining technological specialization, proximity to global leaders, and the structure of knowledge flows. This dual approach, country-by-country and EU-as-a-bloc, allows us to assess whether Europe functions as a coherent technological entity or as a set of heterogeneous national systems.


\section{Descriptive evidence}
\label{sec: descriptive_evidence}
This section present descriptive evidences about several dimensions of the contemporary AI patent ecosystem, revealing a variety of key features, from its geographical and sectoral distribution to the characteristics of the leading players.

\subsection{Geographic concentration}
\begin{figure}[h]
    \begin{center}
    \caption{World 's countries by number of AI patents granted.}
    \label{fig: map}
    \includegraphics[scale=.7]{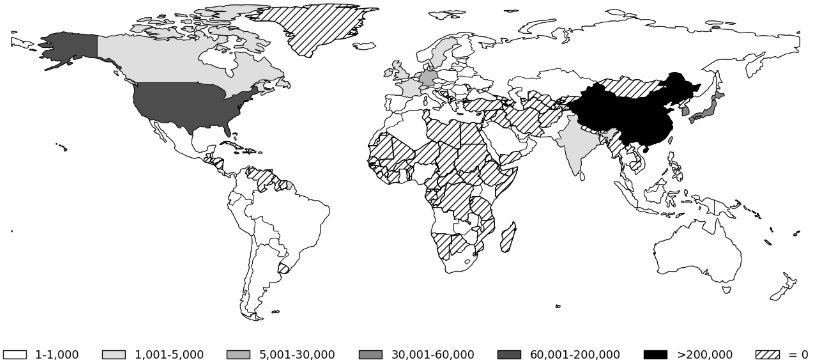}
    \end{center}
    \begin{tablenotes}
\footnotesize 
\item Note: The figure represent world's countries by number of AI patents granted to firms from these states during the sample period 2010-2023.
\end{tablenotes}
\end{figure}

A striking feature of the global AI patent landscape is its high degree of geographic concentration. As shown in Figure \ref{fig: map} China is the undisputed leader, followed by the United States and Japan. Looking at Table \ref{tab: dist_firm_country_I} and \ref{tab: dist_firm_country_II} (Appendix \ref{sec:appendix_tab_graph}), one can see how a small number of countries account for the vast majority of AI patents. Specifically, China hold 220,463 patents filed by entities within its borders, followed by the United States with 80,371 patents. Together, these two countries account for a significant majority of all AI patents in our sample. They are followed at a distance by South Korea (38,054), Japan (36,436) and Germany (6,971). \\

\noindent The European Union is positioned as marginal area for AI patented innovation. In fact, even aggregating all AI patent portfolios in the area, the total count during the sample period amount to 16,689 granted applications, less than Japan, the fourth country in the ranking, despite having a larger number of patentees in AI (2,693 against the 2,092 nipponic ones). This relative underrepresentation reflects Europe’s lower average productivity with a mean number of granted application of around 6 per AI patentee, against higher figures of the major actors, that also show a larger population of AI patenting firms. Compared with the scale-driven patenting strategies of the U.S. and China, Europe’s AI activity remains marginal in volume, and its population of patenting actors smaller. In addition, within its border, the EU shows itself an highly concentrated panorama, with the top five patentees - namely Germany (42\%), France (13\%), The Netherlands (12\%), Sweden (9\%) and Ireland (7\%) - together accounting for almost 83\% of patents production.\\

\noindent This concentration is also evident when considering the nationality of the parent company\footnote{Company with a minimum controlling share of 50.01\%.}, as detailed in Tables \ref{tab: dist_firm_country_guo_I}, and \ref{tab: dist_firm_country_guo_II} (Appendix \ref{sec:appendix_tab_graph}). While the rankings shift slightly, the dominance of China and the United States remains clear. This distinction is important; for instance, a patent filed by a European subsidiary of a U.S. firm would be located in Europe by applicant country but in the U.S. by parent company country. The data shows that AI patenting is not only performed by firms residing in these key countries but is also controlled by parent companies headquartered there. \\

\noindent Finally, although the vast majority of AI-patenting firms are nationally based - 39,174 entities, or about 77\% of the total - multinational corporations exert a disproportionate influence on patenting outcomes. Despite representing only 23\% of active firms (11,505), they account for 230,058 granted AI patents, corresponding to 56\% of the total. By contrast, national companies collectively produced 184,191 patents, or 44\% of the overall output. This asymmetry points to an important distinction: AI patented innovation is numerically driven by national firms, yet its technological frontier is increasingly shaped by the scale, resources, and global reach of multinational players. In other words, while national firms dominate in number, the weight of knowledge production is tilted toward large conglomerates that can leverage extensive R\&D capabilities and cross-border networks.


\subsection{Sectoral distribution}

While AI is often considered a general-purpose technology, its development is currently concentrated in specific sectors. Table \ref{tab: dist_firm_nace} shows the distribution of firms with at least one AI patent across 2-digit NACE industrial sectors. The data clearly indicates that the bulk of AI patenting originates in manufacture (NACE 10-33) and information and communication technology sector (NACE 58-63), immediately followed by professional, scientific and technical activities (NACE 69-75). These three sectors together account for around 67\% of AI-firms, responsible for almost 60\% of AI patents production. An interesting figure coming out of this prospect is the number of applications granted to entities in the education sector (NACE 85). In fact, even if these actors represent less than 3\% of total number of applicants in the AI domain, they account for more than 18\% of the total production, highlighting the central role of universities and research centers in artificial intelligence innovation. \\

\noindent Given the usual large share of manufacturing companies in the population of the majority of developed and developing economies, we reported in Appendix \ref{sec:appendix_tab_graph}, Table \ref{tab: dist_firm_nace_manufacture} the distribution over non aggregated 2-digits NACE sectors of the manufacturing sub-sample. As expected the majority of AI patentees (63\%) are concentrated in computer and electronics (NACE 26), electrical equipment (NACE 27) and machinery and equipment (NACE 28), together accounting for around 68\% of AI patents from manufacturing firms. Interestingly, motorvehicles and trailers (NACE 29), adding up to only 5.2\% of AI patentees in manufacture, are responsible for a share of 16\% of patents in this sub-sample, highlighting the centrality of autonomous driving technologies for the AI innovation domain. Shifting the focus on the European subsample, the distribution of patenting firms closely resemble the global one. However, almost 50\% of AI patents production comes from Manufacturing firms\footnote{Mainly with NACE 26, 27 or 29, alike the global distribution.}, against the global value of $\simeq23\%$. Furthermore, entities from the Education sector account for less than 3\% of the total in Europe, far below the share in the global distribution, suggesting different strategies and commitments for fueling patented AI innovation.\\

\begin{table}[h]
    \centering
        \caption{Distribution of AI-firms and their patent over 2-digits NACE macro sectors.}
        \label{tab: dist_firm_nace}
\resizebox{1\textwidth}{!}{%
    \begin{tabular}{C{2cm} L{8cm} cccc}
    \hline
        NACE & Description & N. firms & \% of total (firms) & N. AI patents & \% of total (AI patents) \\ \hline
        01-03 & Crop, animal production, fishing, hunting, forestry and logging & 135 & 0.3 & 207 & $<$0.1 \\ 
        05-09 & Mining and extraction & 201 & 0.4 & 1,118 & 0.3 \\ 
        10-33 & Manifacture & 12,880 & 25.4 & 94,981 & 22.9 \\ 
        35 & Electricity, gas, steam and air conditioning supply & 682 & 1.3 & 5,298 & 1.3 \\ 
        36-39 & Water, sewerage and waste management & 96 & 0.2 & 163 & $<$0.1 \\ 
        41-43 & Construction, civil engineering and special activities & 1,105 & 2.2 & 2,495 & 0.6 \\ 
        45-47 & Wholesale and retail trade & 3,480 & 6.9 & 16,397 & 4.0 \\ 
        49-53 & Transport and storage & 434 & 0.9 & 1,116 & 0.3 \\ 
        55-56 & Accommodation  and food service activities & 70 & 0.1 & 126 & $<$0.1 \\ 
        58-63 & Information and communication & 12,303 & 24.3 & 106,658 & 25.7 \\ 
        64-69 & Financial, insurance and real estate activities & 1,279 & 2.5 & 11,827 & 2.9 \\ 
        69-75 & Professional, scientific and technical activities & 8,693 & 17.2 & 43,989 & 10.6 \\ 
        77-79 & Administrative and support service activities & 226 & 0.4 & 717 & 0.2 \\ 
        80-82 & Security and investigation activities & 876 & 1.7 & 8,851 & 2.1 \\ 
        84 & Public administration and defence & 153 & 0.3 & 1,417 & 0.3 \\ 
        85 & Education & 1,474 & 2.9 & 75,658 & 18.3 \\ 
        86-88 & Human health and social work activities & 1,014 & 2.0 & 4,209 & 1.0 \\ 
        90-93 & Arts, entertainment and recreation & 141 & 0.3 & 564 & 0.1 \\ 
        94-96 & Other services activities & 209 & 0.4 & 1,171 & 0.3 \\ 
        97-98 & Activities of households as employers; undifferentiated goods & 3 & $<$0.1 & 4 & $<$0.1 \\ 
        Missing & - & 5,227 & 10.3 & 37,283 & 9.0 \\ 
\hline \hline
\end{tabular}
}
\begin{tablenotes}
\footnotesize 
\item Note: The table reports the distribution of firms with at least one granted AI-patent, and their respective patents production, over 2-digits NACE sectors. Period 2010-2023.
\end{tablenotes}
\end{table}

\noindent Despite this concentration, the data also reveals a broad diffusion of AI patenting across nearly all sectors of the economy, from agriculture (NACE 01-03; 135 firms) financial, insurance and real estate (NACE 64-69; 1,279 firms) and human health and social work activities (NACE 86-88; 1,014 firms). This suggests that while core AI development is driven by the tech industry, its applications are being explored and integrated across the entire economic landscape.\\

\begin{figure}[h]
    \begin{center}
    \caption{AI patents granted to applicants in top 4 NACE sectors by number of inventions.}
    \label{fig: top_4_nace}
    \includegraphics[scale=.7]{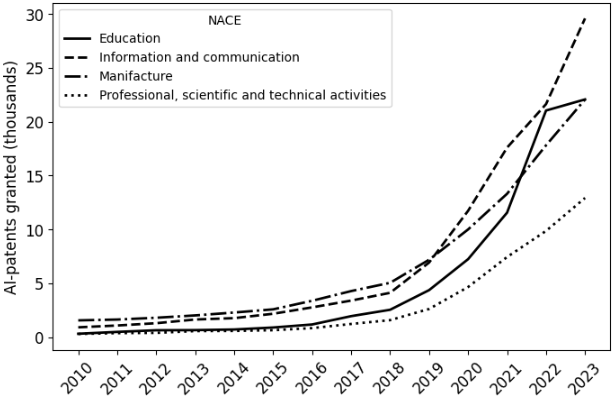}
    \end{center}
    \begin{tablenotes}
\footnotesize 
\item Note: The figure represent the distribution over time of AI-grants of patentees in the top 4 NACE Rev.2 sectors, by number of granted patents. Period 2010-2023.
\end{tablenotes}
\end{figure}

\noindent When examining the distribution of AI-related grants across the main four NACE sectors (accounting for around 77\% of total patented inventions) over time (Figure \ref{fig: top_4_nace}), a clear shift in sectoral leadership emerges. Manufacturing firms initially dominated in AI patent activities, consistently holding the leading position until 2019. From that point onward, the information and communication (ICT) sector became the most prominent domain for AI granted applications. A further notable shift occurred in 2021, when the education sector overtook manufacturing and, by the following year, nearly equaled the ICT sector in terms of AI patent production. \\

\noindent These dynamics align with broader developments in the field of artificial intelligence. In earlier years, simpler machine learning models - such as those used in industrial applications for tasks like product sorting or fault detection in production lines - were already mature and effective. However, the emergence of more sophisticated architectures, notably transformer-based models, along with the increasing availability of high-quality data and rapid advances in computational power, has enabled a broader range of applications. These technological advancements have not only enhanced the performance of legacy systems but also expanded AI’s applicability across sectors, contributing to the observed shifts in patenting leadership.

\noindent It is important to point out a caveat of Figure \ref{fig: top_4_nace} concerning the interpretation of sectoral leadership in AI patenting. While the NACE classification allows us to assign patents to broad industries, it does not reveal the intended use of the underlying inventions. For instance, it remains unclear whether manufacturing firms patent AI technologies primarily for their own production processes, or whether ICT and professional service firms develop AI solutions that are subsequently deployed in manufacturing and other sectors. The relative proportions of “self-use” versus “external supply” across industries cannot be identified from the data, and this limits the extent to which we can infer the underlying division of labor in AI innovation from patenting trends alone. 

\subsubsection{Market concentration}
To assess the degree of market concentration in AI patenting across industries, we compute the \textit{$CR_{q}$ concentration ratio}, a standard measure of market dominance. For each two-digit NACE sector $s$, let $AIP_{is}$ denote the number of AI patents granted to firm $i$ operating in sector $s$. The total number of AI patents in the sector is then:

\begin{equation}
    P_{s} = \sum_{i=1}^{N_s} P_{is}
\end{equation}

\noindent where $N_s$ is the number of firms in sector $s$ with at least one granted AI patent. The $CR_{q}$ concentration ratio is defined as the share of AI patents accounted for by the five most prolific firms within each sector:

\begin{equation}
    CR_s = \frac{\sum_{i \in \text{Top\;q}_s} P_{is}}{P_s}
\end{equation}

\noindent The index ranges from $0$ (perfectly diffuse patenting activity) to $1$ (complete monopolization by five firms). Values closer to one indicate a higher degree of concentration and, hence, a smaller number of dominant patent holders in that sector. \\

\begin{figure}[h]
    \centering
    \caption{Kernel density of the $CR_5$ concentration ratio across NACE sectors.}
    \includegraphics[width=0.7\textwidth]{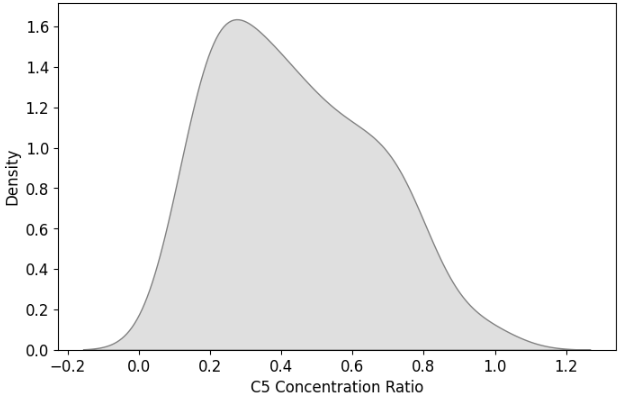}
    \label{fig:c5_density}
    \begin{tablenotes}
\footnotesize 
\item Note: The figure represent the density distribution of the $CR_5$ index across 2-digits NACE. Period 2010-2023.
\end{tablenotes}\end{figure}

\noindent Figure \ref{fig:c5_density} presents the kernel density of the $CR_{5}$ ratio across all NACE sectors. The distribution is centered around $CR_{5} \approx 0.25{-}0.30$, indicating that, in the median sector, the five leading firms collectively account for roughly one quarter to one third of total AI patenting activity. This suggests a moderate degree of concentration, where innovative capacity is shared among several players rather than monopolized by a few dominant ones. At the same time, the long right tail of the distribution-extending toward values above $0.8$-reveals that a subset of sectors exhibits highly concentrated AI patent production. These likely correspond to technologically intensive or
capital-heavy industries, such as semiconductors, defense, or specialized manufacturing, where entry barriers are high and innovation is led by a handful of large incumbents. The absence of density near zero further indicates that no sector exhibits completely diffuse AI patenting. In other words, AI innovation appears structurally concentrated across the economy, reflecting both the scale requirements of frontier research and the uneven distribution of data, computational resources, and research capabilities across firms. Overall, the observed pattern points to a hybrid structure: while AI has diffused beyond core ICT and manufacturing sectors, its production remains dominated by a small set of technologically advanced firms within each industry.

\subsection{Key players and firm characteristics}

An examination of the leading patent applicants (Table \ref{tab: top_innovators}) reveals that the AI patent landscape is dominated by a mix of large, established technology corporations and major public research institutions, particularly from China. International Business Machines Corp. (IBM) leads the corporate world with 9,138 patents, followed by Chinese tech giants Tencent (4,580) and Baidu (4,372), and South Korea's Samsung (3,737). American firms Google and Microsoft also feature prominently, with respectively 3,331 and 2,863 granted application related to AI technologies. Notably, the list of top applicants is heavily populated by Chinese universities, such as Zhejiang University and the University of electronic science and technology of China, highlighting the significant role of public research in China's national AI strategy. Not a single European company appears among the top position of the global ranking, the bigger player in terms of granted AI patents during the sample period is fact the Dutch Koninklijke Philips, in 60th position, with 956 granted applications. It follow as 66th the German Robert Bosch (858) and as 98th the Irish Accenture (616).\\

\begin{table}[h]
    \centering
        \caption{Top 20 patentees by n.of AI patents}
        \label{tab: top_innovators}
\resizebox{1\textwidth}{!}{%
    \begin{tabular}{L{9cm} C{2.5cm} C{3cm} C{2.5cm} L{4.5cm} C{3cm} C{2cm}}
    \hline
        Company name & Year of incorporation & Country & NACE & NACE Description & Size & N. AI patents \\ \hline
    International Business Machines Corp. & 1911 & United States of America & 58-63 & Information and communication & Very large company & 9,138 \\ 
    Tencent Technology (Shenzhen) Co., Ltd. & 2000 & China & 58-63 & Information and communication & Very large company & 4,580 \\ 
    Beijing Baidu Netcom Science And Technology Co., Ltd. & 2001 & China & 58-63 & Information and communication & Very large company & 4,372 \\ 
    Samsung Electronics Co., Ltd. & 1969 & South Korea & 10-33 & Manufacture & Very large company & 3,737 \\ 
    Google Llc & 1998 & United States of America & 58-63 & Information and communication & Very large company & 3,331 \\ 
    Microsoft Technology Licensing, Llc & 1975 & United States of America & 58-63 & Information and communication & Very large company & 2,863 \\ 
    Zhejiang University & 2001 & China & 85 & Education & Very large company & 2,725 \\ 
    University Of Electronic Science And Technology Of China & 1956 & China & 85 & Education & Small company & 2,672 \\ 
    Nec Corporation & 1899 & Japan & 58-63 & Information and communication & Very large company & 2,444 \\ 
    Ping An Technology (Shenzhen) Co., Ltd. & 2008 & China & 58-63 & Information and communication & Very large company & 2,177 \\ 
    Tsinghua University & 1911 & China & 85 & Education & Very large company & 2,121 \\ 
    State Grid Corporation Of China & 2003 & China & - & - & Small company & 2,075 \\ 
    Xi'an Electronic Technology University & 1931 & China & 85 & Education & Small company & 2,068 \\ 
    Fujitsu Limited & 1935 & Japan & 58-63 & Information and communication & Very large company & 2,013 \\ 
    Amazon.com, Inc. & 1996 & United States of America & 45-47 & Wholesale and retail trade & Very large company & 1,947 \\ 
    Baidu Online Network Technology (Beijing) Co., Ltd. & 2000 & China & 58-63 & Information and communication & Very large company & 1,940 \\ 
    Beijing Aerospace University & 1952 & China & 85 & Education & Very large company & 1,887 \\ 
    South China University Of Technology & 1952 & China & 85 & Education & Very large company & 1,845 \\ 
    Huawei Technologies Co., Ltd. & 1987 & China & 10-33 & Manufacture & Very large company & 1,792 \\ 
    Mitsubishi Electric Corporation & 1921 & Japan & 10-33 & Manufacture & Very large company & 1,679 \\
\hline \hline
\end{tabular}
}
\begin{tablenotes}
\footnotesize 
\item Note: The table reports applicant-level information about the top 20 applicants by the number of granted AI patents. Period 2010-2023.
\end{tablenotes}
\end{table}

\begin{figure}[h]
    \begin{center}
    \caption{AI applications granted to top 5 applicants by patent portfolio.}
    \label{fig: top_5_applicants}
    \includegraphics[scale=.65]{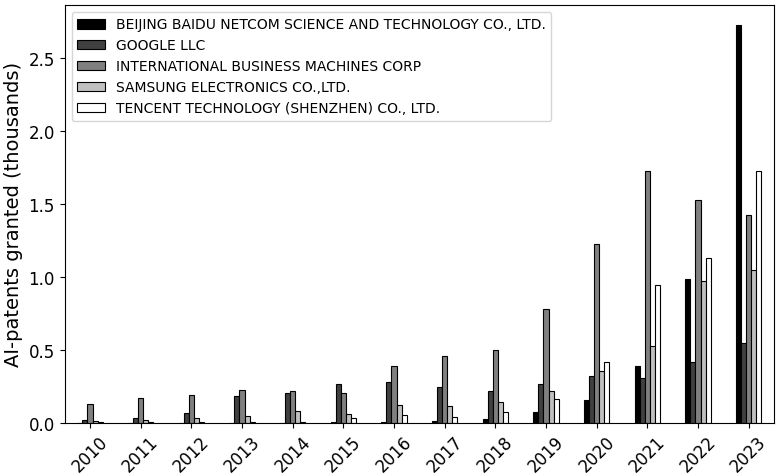}
    \end{center}
    \begin{tablenotes}
\footnotesize 
\item Note: The figure represent the distribution over time of AI-grants for the top 5 applicants by size of AI-portfolio. Period 2010-2023.
\end{tablenotes}
\end{figure}

\noindent When analyzing the temporal dynamics of AI patent activity among the top five entities (Figure \ref{fig: top_5_applicants}), the clear competition between leading U.S. and Chinese innovators is further highlighted. IBM and Google historically dominant forces in AI patent filings, maintained a consistent presence over the years. However, their long-standing lead was overtaken in 2023 by the Chinese company Baidu and its domestic competitor Tencent, which ascended to the top of the rankings in just two years. 
The global and domestic presence of both Baidu and Tencent - particularly their increasingly horizontal expansion across markets and technologies - invites a dual interpretation of this surge. On one hand, it may signals intensified R\&D efforts, supported by access to large and diverse datasets. On the other, it may reflect a more distributed and sectorally diverse production of AI patented technologies across technological domains, particularly within the Chinese patent ecosystem. \\

\noindent Another interesting characteristic of AI-innovating firms is their age distribution\footnote{Please refer to Figure \ref{fig: age_dist} in Appendix \ref{sec:appendix_tab_graph} for the full distribution.}. The mean is centered around 15-20 years of age, suggesting a strong presence of young firms, likely startups founded during the recent AI applications surge. This reflects the classic Schumpeterian dynamic of 'creative destruction,' where new firms are a primary engine of radical innovation - consider for example the impact of models from OpenAI, a firm founded in 2015, now seen as a main competitor of tech giants like Google and Amazon. This trend is particularly visible in countries with a high number of new entrants, such as China, South Korea, and Israel, where the mean age of AI-patenting firms is just above 15 years (Table \ref{tab: dist_firm_country_I} in Appendix \ref{sec:appendix_tab_graph}). In the right tail of the distribution we find older, multinationals like IBM (founded in 1911) and Mitsubishi (incorporated in 1921) and other established industrial giants, together with historical research institutions like Tsinghua University (1911) or Beijing Aerospace University (1952). This is a characteristic of innovation systems in countries like Japan where the mean firm age is 48 years, Germany where is 40 years and Europe (taken as a whole) and the United States, with a 31 years mean, reflecting the significant role of long-established corporations that have successfully adapted their R\&D focus to include AI. These countries however, show a distribution of the age of AI innovators more dispersed, with larger standard deviation, suggesting an heterogeneous environment where young, agile firms and startups innovate alongside deeply entrenched technology leaders.


\subsection{ Technological proximity}
\label{sec: tech_prox}
A further descriptive dimension concerns the technological proximity of AI patent portfolios across countries. By comparing the overlap of technological classes, we can assess whether national innovation systems are converging on common technological trajectories or remain differentiated. This measure complements the geographic, sectoral, and organizational evidence presented above by highlighting similarities and divergences in the composition of AI innovation. In order to do so we leverage the min-complement proximity measure proposed by \citet{bar_measure_2012}. This measure captures the share of overlapping AI inventions between two firms' patent portfolios, with values ranging from 0 (completely distinct technological profiles) to 1 (identical technological profiles).

\noindent For any two firms $i$ and $j$, let:
\begin{equation}
\mathbf{P}_i = (p_{i1}, p_{i2}, \ldots, p_{ik}, \ldots, p_{in}) \text{ and } \mathbf{P}_j = (p_{j1}, p_{j2}, \ldots, p_{jk}, \ldots, p_{jn})
\end{equation}
represent their respective patent portfolio vectors, where $p_{ik}$ denotes the share of firm $i$ AI patent classes in the 4-digits CPC/IPC technology class $k$ out of $n$ relevant, with $\sum_{k=1}^{n} p_{ik} = 1$. We focus exclusively on AI patent portfolios rather than firms' entire patent holdings to ensure that our proximity measure reflects genuine technological relatedness within the artificial intelligence domain, avoiding dilution from unrelated technological activities. The technological proximity between firms $i$ and $j$ is then defined as:
\begin{equation}
\text{Proximity}(\mathbf{P}_i, \mathbf{P}_j) = \sum_{k=1}^{n} \min\{p_{ik}, p_{jk}\}
\end{equation} \\

\noindent Table \ref{tab: proximity_stat} reports descriptive statistics for the technological proximity measure computed on our sample of AI patent portfolios. Notably, the average and median proximity level across all country pairs are close and just above 50\%, indicating substantial heterogeneity in AI specialization patterns across countries. This finding could suggests that the observed clustering of patent citations within national borders may be partially explained by differences in technological focus rather than purely geographic or institutional factors. However, this hypothesis is challenged by the remarkably high proximity (i.e. 0.78) between the U.S. and Chinese AI portfolios. Despite these two countries representing the dominant poles of AI patented innovation and exhibiting strong home bias in citation patterns, their AI patent distributions are highly similar. This suggests that citation clustering reflects stylized facts found by \cite{jaffe_geographic_1993} about patent citation behavior - such as language barriers, institutional familiarity, and network effects - rather than technological specialization in AI, despite challenging results of \cite{thompson_patent_2005}. \\

\begin{table}[h]
    \centering
        \caption{Descriptive statistics on proximity distribution.}
        \label{tab: proximity_stat}
    \begin{tabular}{ccccc}
    \hline
        Mea& Median&	Std.dev.&	Max &	Min \\ \hline
        0.34&	0.33&	0.21&	0.88&	0 \\
\hline \hline
\end{tabular}
\begin{tablenotes}
\footnotesize 
\item Note: The table reports descriptive statistics about the distribution of the proximity measure in our sample.
\end{tablenotes}
\end{table}

\begin{figure}[h]
    \begin{center}
    \caption{AI technological proximity to the U.S. and China.}
    \label{fig: proximity_sides}
    \includegraphics[scale=.85]{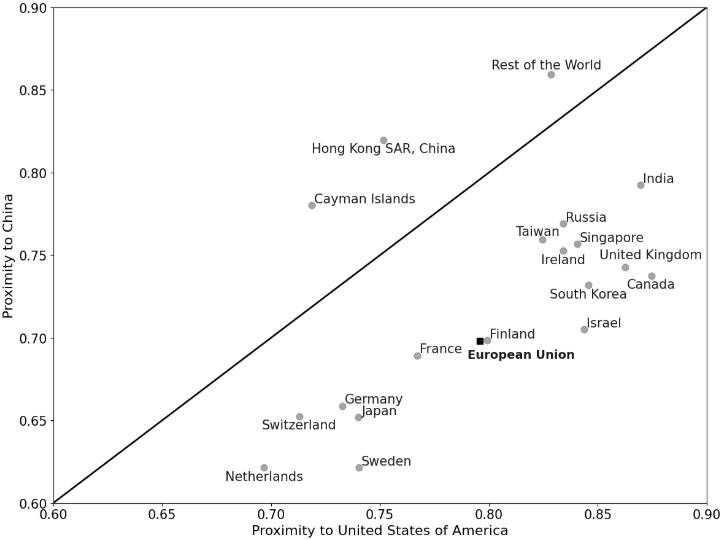}
    \end{center}
    \begin{tablenotes}
\footnotesize 
\item Note: The figure represent AI portfolios technological proximity to the U.S. and China. In the category \textit{Rest of the World} are aggregated countries which total AI patents production sum accounts for less than 1\% of the total. The \textit{European Union} value report proximity computed by aggregating EU countries portfolios. Period 2010-2023.
\end{tablenotes}
\end{figure}

\noindent Given the bipolar structure of global AI patenting centered on the U.S. and China, we examine the relative technological proximity of other countries to these two leaders. Figure \ref{fig: proximity_sides} presents the technological proximity to U.S. and Chinese AI portfolios for the top 20 countries by AI patent volume (collectively accounting for approximately 99\% of global AI patent production).\footnote{See Appendix \ref{sec:appendix_tab_graph}, Figure \ref{fig: proximity_heatmap} for the complete technological proximity heatmap.} A striking pattern emerges: while the overall average proximity is around 34\%, all top 20 countries (U.S. and China included) exhibit proximity values exceeding 60\% to both the U.S. and China, with most countries aligning more closely with the U.S. portfolio than with China's. The notable exceptions\footnote{Also the residual category "Rest of the World" exhibits a closer proximity to China, but it does account for less than 1\% to total patents production.} are Hong Kong and the Cayman Islands, which show stronger proximity to China. \\

\noindent These findings reveal a core-periphery structure in global AI patenting systems with several important implications. First, the high proximity of all major AI-patenting countries to both the U.S. and China, contrasted with the low overall average, suggests that these two leaders define a common technological core that other countries follow. This creates a "convergence to the leaders" dynamic where successful AI patented innovation requires alignment with the technological trajectories established by the U.S. and China. Second, despite the competitive rhetoric surrounding U.S.-China technological rivalry, both countries are pursuing remarkably similar AI research directions, potentially indicating that certain AI technological pathways are dominant regardless of national innovation systems.
Finally, the slightly stronger alignment with U.S. patterns (except for Hong Kong and Cayman Islands) may reflect the global influence of U.S. technology platforms, research institutions, and technical standards, suggesting that the U.S. maintains a subtle but persistent advantage in shaping global AI development trajectories despite China's rapid rise in patent volume. \\

\noindent When we aggregate AI patent portfolios of companies in the European Union we find that the average technological proximity with the two major patentees countries are close to the ones of France and Finland: around 80\% to the U.S. and 70\% to China. This pattern suggests that, despite the global diffusion of AI technologies, the European AI innovation landscape remains embedded within a transatlantic technological sphere, likely reflecting shared industrial linkages, research collaborations, and institutional frameworks that favor compatibility with the U.S. innovation system. Nevertheless, the EU’s distance from the diagonal is not substantial, implying that its AI patent portfolio also exhibits a degree of convergence with China’s, consistent with the increasing internationalization of AI research and the global circulation of technological knowledge. This situation underscores the absence of an autonomous European core: the old continent remains positioned within the western innovation orbit rather than forming a third global pole. Finally, EU countries are relatively scattered around the plane, and with respect to the aggregated patent portfolio, suggesting no particular correlation between patented AI innovation trajectories.


\section{Comparative advantage}
\label{sec: rca}

To assess the relative specialization of countries in artificial intelligence technologies, it is crucial to go beyond absolute measures such as patent counts and consider each country's position in the global technological landscape. For this purpose, we leverage the Revealed Comparative Advantage (RCA) index, originally proposed by \cite{balassa_trade_1965}, an index used in international economics that measures whether a country has a relative advantage or disadvantage in producing and exporting a particular category of goods or services, based on observed trade patterns. For our application, we adapted the RCA of a country $i$ in AI patents defining it as:
\[
\text{RCA}_{i}^{AI} = \frac{\left(\frac{P_{i}^{AI}}{P_{i}^{\text{Total}}}\right)}{\left(\frac{P_{\text{World}}^{AI}}{P_{\text{World}}^{\text{Total}}}\right)}
\]
where $P_{i}^{AI}$ denotes the number of AI patents granted to applicants from country $i$, $P_{i}^{\text{Total}}$ is the total number of patents (across all technological fields) granted to that country\footnote{In order to maintain consistency with applicants country information source, we fetched these figures from ORBIS IP.}, $P_{\text{World}}^{AI}$ is the total number of AI patents granted globally, and $P_{\text{World}}^{\text{Total}}$ is the total number of patents granted worldwide. An RCA value greater than one ($\text{RCA}_{i}^{AI} > 1$) indicates that country $i$ is relatively more specialized in AI technologies than the world average, suggesting a revealed comparative advantage in this domain. \\

\noindent Using the RCA index in the context of AI patenting provides a normalized and comparative perspective on technological specialization. It allows us to identify countries that, despite possibly producing a low absolute number of patents, are highly focused on artificial intelligence relative to their overall patenting activity. Conversely, countries with large patent volumes but low RCA scores may appear less specialized. Therefore, the RCA measure complements absolute indicators by capturing the intensity of focus or prioritization on AI within each country’s patent portfolio. This is particularly relevant for assessing strategic national priorities in emerging technologies, informing technology policy, and understanding global competitive dynamics in AI patenting activities. Table \ref{tab: top_15_rca} report the RCA index together with percentage shares of AI patents over total for the top 15\footnote{Please refer to Appendix \ref{sec:appendix_tab_graph}, Tables \ref{tab: rca_high} and \ref{tab: rca_low} for the full set of figures about RCA index.} countries by number of granted applications in our sample period, in addition results aggregate for EU countries. \\

\begin{table}[h]
    \centering
        \caption{RCA index - Top 15 countries by AI patents and E.U.}
        \label{tab: top_15_rca}
\resizebox{.9\textwidth}{!}{%
    \begin{tabular}{lcccc}
    \hline
        Country & N. AI patents & N.Patents & \% of total & RCA \\ \hline
        China & 220,463 & 14,653,708 & 2 & 0.47 \\ 
        United States of America & 80,371 & 933,528 & 9 & 5.15 \\ 
        South Korea & 38,054 & 965,745 & 4 & 2.09 \\ 
        Japan & 36,436 & 2,112,676 & 2 & 0.86 \\ 
        \textit{European Union}	& \textit{16,689} & \textit{973,110} & \textit{2} & \textit{0.86} \\
        Germany & 6,971 & 459,585 & 2 & 0.76 \\ 
        Taiwan & 6,586 & 265,484 & 2 & 1.26 \\ 
        United Kingdom & 2,517 & 86,009 & 3 & 1.49 \\ 
        France & 2,143 & 138,108 & 2 & 0.78 \\ 
        Netherlands & 1,950 & 79,111 & 2 & 1.25 \\ 
        Canada & 1,771 & 49,634 & 4 & 1.81 \\ 
        Cayman Islands & 1,753 & 20,419 & 9 & 4.36 \\ 
        Israel & 1,609 & 20,579 & 8 & 3.97 \\ 
        Sweden & 1,439 & 53,020 & 3 & 1.38 \\ 
        India & 1,255 & 15,256 & 8 & 4.18 \\ 
        Ireland & 1,186 & 13,311 & 9 & 4.52 \\ 
\hline \hline
\end{tabular}
}
\begin{tablenotes}
\footnotesize 
\item Note: The table reports data on patents and AI-RCA index for the top 15 countries by number of granted AI applications in addition to aggregated data for the European Union. Period 2010-2023.
\end{tablenotes}
\end{table}

\noindent Although China, Japan, Germany, and France rank among the top countries globally in terms of the absolute number of granted AI patents, their RCA values fall below one. When aggregating patents portfolios of European country the situation is analogue, with a RCA index equal to Japan: 0.86. This outcome reflects the fact that, relative to their overall patenting activity, these countries are not disproportionately focused on AI technologies. Instead, their patenting efforts are broadly distributed across a wide range of technological domains. A low RCA, therefore, does not imply a lack of technological capacity or engagement, but rather indicates that artificial intelligence represents a smaller share of the country's patent specialization when compared to the global average. This distinction highlights the analytical value of the RCA index, which allows for a more nuanced understanding of national patenting priorities.
On the other hand U.S., Cayman Islands\footnote{Given well established fiscal advantages, data related to this country are likely biased on a geographical level.}, Qatar, Israel, and Ireland show the largest\footnote{Together with India, however this country has been excluded from our comments due to the known lack of coverage of our data source.} values of the RCA index, indicating that even if their final output is lower in absolute terms, then the one in different geographical areas, their relative specialization is far more intense. This aspect likely reflect different strategies in R\&D investments, where not only artificial intelligence is seen as a major area of interest but where maybe other kind of patent domains incorporate AI technologies in an horizontal way. \\

\noindent However, when we look at the number of CPCs classes in which AI patenting occur, they seems to have a positive and exponential relationship with the number of granted applications by country (Figure \ref{fig: cpc_scatter} in Appendix \ref{sec:appendix_tab_graph}). In fact the ranking of countries "touching" the largest number of technological domains, closely resemble the one of the total number of patents. This suggest that the intensity of patenting production, rather than the scope of AI granted applications is relatively higher in country with a value of the RCA index larger than 1. \\


\section{Impact and spillover}
\label{sec: impact}

While the absolute number of granted applications provides valuable insights into patents output, it is not enough to understand the broader impact and influence of AI technologies. Since patent citations potentially serve as indicators of patent quality - see e.g. \cite{albert_direct_1991, trajtenberg_university_1997} and knowledge spillovers, we analyze applications bibliographic footprint through citation patterns. This section analyzes citations\footnote{We examine forward patent citations (i.e. number of patents that cite an AI patent in its references) at the family level.} patterns to identify signals of patent quality, impact, and the geographical and sectoral dynamics of knowledge flows in the AI patenting ecosystem. \\ 

\noindent However, it is important to acknowledge that patent quality measurement is inherently complex and multidimensional. Recent research by \cite{higham_patent_2021} demonstrates that commonly used patent quality indicators - including forward citations, patent renewals, and stock market reactions to patent grants - show generally poor agreement on which patents constitute "high-quality." Their findings suggests that because 'quality' is an intrinsically multidimensional concept, there cannot be a single best metric. Given these limitations, our citation analysis should be interpreted as capturing specific dimensions of patent influence and knowledge flows rather than providing a comprehensive measure of overall patent quality. The technology-specific nature of these relationships is particularly relevant for AI patents, which span diverse technological domains and application areas.

\subsection{Geographic patterns of citation}

Our analysis reveals a striking divergence between patent production volume and (forward) citation impact across countries\footnote{Please refer to Tables \ref{tab: citations_dist_p1} and \ref{tab: citations_dist_p2} in Appendix \ref{sec:appendix_tab_graph} for the full distribution at a country level.}. The United States emerges as the dominant force in terms of forward citation influence, accounting for 1,147,971 citations - nearly 47\% of the total citations received by AI patent families during our sample period. This is followed by China with 651,202 citations ($\simeq27\%$) and Japan with 212,548 citations ($\simeq9\%$). \\

\noindent The disparity between China’s patent production leadership and its relatively lower citation impact presents a compelling case study. Despite producing significantly more AI patents than any other country, China’s citation performance suggests a lower degree of technological disruptiveness and influence compared to U.S. granted applications. This pattern may reflect several underlying factors. First, the rapid expansion of China’s
AI patent portfolio may include a higher proportion of incremental technologies rather than breakthrough ones that typically attract more citations - a situation operationally similar to the so called \textit{evergreen} patenting strategies in the pharmaceutical sector \citep{kesselheim_biomedical_2006}. Second, the timing of patent grants may influence citation patterns, as newer patents naturally have less time to accumulate citations. Third, differences in patent quality assessment and strategic patenting behaviors between countries may contribute to varying citation intensities. \\

\begin{table}[h]
    \centering
        \caption{Forward citations by family}
        \label{tab: top_5_citation_matrix}
\resizebox{.9\textwidth}{!}{%
    \begin{tabular}{rccccc} \hline ~ & ~ & ~ & \textit{Cited} ~ & ~ & \\
        \multicolumn{1}{r}{\textit{Citing}} & U.S. & China & Japan & EU & South Korea \\ 
        \cline{2-6}
        U.S. & \textbf{759,316} & 26,331 & 64,913 & 64,807 & 32,998 \\ 
        China & 137,950 & \textbf{590,617} & 32,750 & 21,467 & 27,318 \\ 
        Japan & 51,807 & 6,519 & \textbf{77,276} & 8,425 & 6,463 \\ 
        EU & 55,021 & 4,921 & 10,112 & \textbf{19,498} & 3,833 \\ 
        South Korea & 37,923 & 5,781 & 13,168 & 4,864 & \textbf{48,517} \\ 
\hline \hline
\end{tabular}
}
\begin{tablenotes}
\footnotesize 
\item Note: The table reports citations at a family level for AI patents. Countries on rows cite countries on columns. Period 2010-2023.
\end{tablenotes}
\end{table}

\noindent Given that the U.S., China, Japan, Europe and South Korea account for roughly 93\% of total AI citations\footnote{The distribution of forward citations of AI patents is in fact strongly right skewed - refer to Table \ref{tab: citation_stat} in Appendix \ref{sec:appendix_tab_graph} for summary statistics at a patent family and country level.}, we focused on source and destination of references for this specific subset. The citation matrix in Table \ref{tab: top_5_citation_matrix} reveals pronounced patterns of self-citation and regional clustering. The United States, China, and Japan demonstrate strong tendencies toward self-citation, with domestic patents citing other domestic applications at significantly high rates - coherently with \cite{jaffe_geographic_1993} findings. This geographic clustering of citations could reflects several phenomena. First, universities, research institutes, and companies within the same country often maintain closer collaborative relationships, leading to more frequent knowledge exchange and citation. Second, patents targeting similar markets or regulatory environments are more likely to reference each other, creating geographic citation clusters. On the other hand, European firms exhibit distinctive citation behavior, directing a majority of their citations toward the U.S. applications rather than domestic patents. This pattern suggests EU’s position as a technology adopter and adapter in AI, leveraging foundational patents developed elsewhere while contributing its own specialized applications. Similarly, South Korean firms demonstrate nearly equal rates of self-citation and U.S. citation, indicating a balanced approach between domestic innovation building and international knowledge absorption. \\

\begin{figure}[h]
    \begin{center}
    \caption{Citations network in AI patents by economic region.}
    \label{fig: cit_flows}
    \includegraphics[scale=.68]{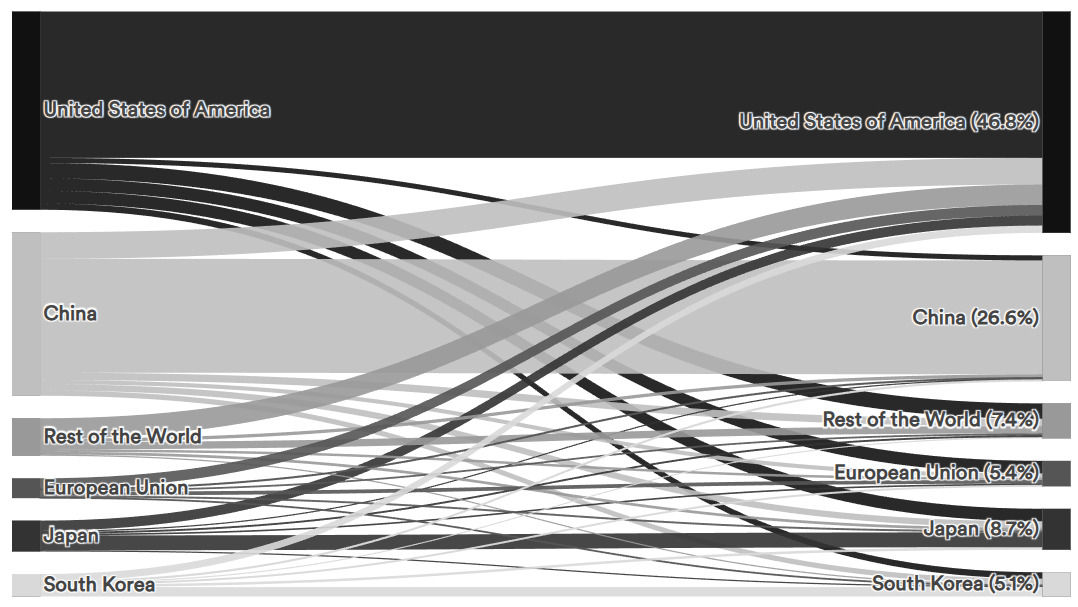}
    \end{center}
    \begin{tablenotes}
\footnotesize 
\item Note: The figure represent forward citations flows of AI patents. Citing countries (any patent) on the left, cited (AI grants) on the right. Percentages are with respect of total AI forward citations.    Period 2010-2023.
\end{tablenotes}
\end{figure}

\noindent This landscape of AI citation is confirmed also at a global level, when observing the network of forward citations flows in Figure \ref{fig: cit_flows}. The network analysis illustrates a bipolar structure of global AI patenting production, with the United States and China forming the two dominant poles of citation activity ($\simeq 73\%$ of the total). The majority of citations within these regions are directed toward granted applications from the same geographic area, with relatively weaker but still significant cross-regional flows. This pattern suggests the emergence of regional patent ecosystems with distinct technological trajectories and knowledge bases, while maintaining connections that facilitate global knowledge diffusion. The relatively limited citation flows to other regions highlight the concentration of AI patented technologies leadership and suggest potential barriers to knowledge transfer, including language differences, institutional variations, and technological focus areas. However, the U.S., Japan, Europe and South Korea, also show a percentage of foreign citations of around 30\%, marking their patents are more internationally relevant than Chinese ones. Furthermore, despite having less than an half of granted AI applications than South Korea, EU shows roughly the same number of citations ($\simeq 5\%$ of the total), signaling possible higher quality of applications, or an innovation strategy of the Asian country more aligned with the Chinese one, perhaps centered on marginal innovations patenting.

\subsection{Survival analysis of citation dynamics}
Recent empirical evidence suggests that the speed at which patents receive their first citation serves as a meaningful signal of patent value and technological impact \citep{gay_determinants_2005, marco_dynamics_2007, fisch_value_2017}. Building on these insights, we examine the temporal dynamics of AI patent citations using survival analysis to assess not only whether patents are cited, but how quickly they enter the knowledge diffusion process - a dimension that may reveal differences in the technological importance and influence of AI inventions across countries.
To complement the analysis of citation counts and spillovers, we estimate the probability that a patent remains uncited over time using a non-parametric survival model. Let $T$ denote the time (in months) between the grant of a patent and its first citation. The survival function is defined as:

\[
S(t) = P(T > t),
\]

\noindent representing the probability that a patent has not been cited by time $t$. Following multiple similar applications\footnote{See for example \cite{xie_survival_2011} and \cite{fisch_value_2017}.} in the literature, we estimate $S(t)$ using the Kaplan–Meier estimator:

\[
\hat{S}(t) = \prod_{t_i \leq t} \left( 1 - \frac{d_i}{n_i} \right),
\]

\noindent where $t_i$ is the $i$-th distinct citation time, $d_i$ is the number of patents cited at $t_i$, and $n_i$ is the number of patents still at risk (i.e., not yet cited) immediately before $t_i$. 
This estimator accommodates right-censoring, as many patents have not yet received citations by the end of the observation window. We compute $\hat{S}(t)$ separately for each major\footnote{Countries excluded from the graph together account for around 7\% of total citations, making their representation not informative and noisy for other curves.} patenting country to visualize and compare their citation lag distributions. \\

\begin{figure}[h]
    \begin{center}
    \caption{Estimated Kaplan–Meier survival curves for AI patents forward citations.}
    \label{fig: survival_curve}
    \includegraphics[scale=.8]{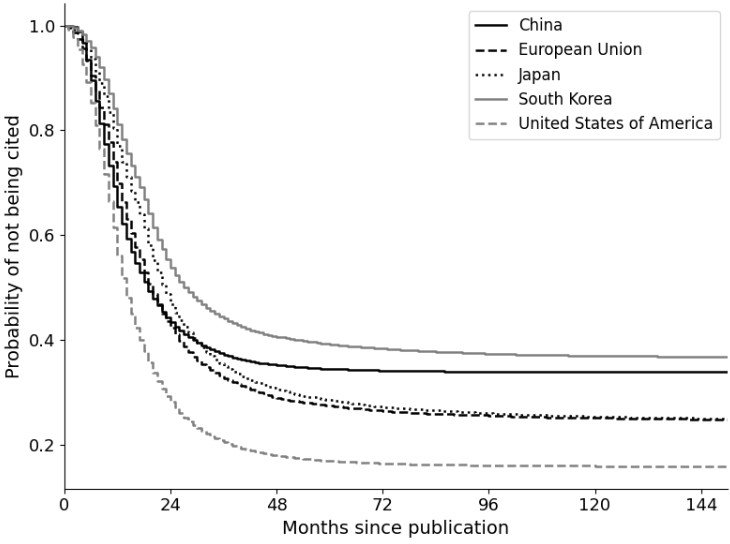}
    \end{center}
    \begin{tablenotes}
\footnotesize 
\item Note: Estimated Kaplan–Meier survival curves for the main AI patenting economies. The vertical axis represents the probability that a patent remains uncited, while the horizontal axis measures the number of months since earliest publication.
\end{tablenotes}
\end{figure}

\noindent Figure \ref{fig: survival_curve} reports the estimated Kaplan–Meier survival curves for the main AI patenting economies. Several important patterns emerge: first, citation delays differ markedly across countries. 
AI patents from the United States exhibit the steepest decline in $S(t)$, indicating that they are cited sooner and more frequently than those from other regions. This finding is consistent with our earlier citation analysis, which identified the United States as the main source of influential AI patents. Second, Japanese and European patents display intermediate survival rates, suggesting a slower diffusion of their technological influence. Chinese patents share the former two curves steepness up to 2 years after publication, then experience flat dynamics. South Korean patents are characterized by even flatter curves, implying longer lags before first citation and lower international visibility. Finally, after approximately 120 months, all survival functions tend to flatten, indicating that most citation activity occurs within the first ten years after patent publication. \\

\noindent Beyond the steepness of the survival curves, it is also important to comment on their flattening points. While approximately 15\% of U.S. patents remain uncited, China and South Korea exhibit the highest shares of uncited patents during the sample period-around 37\% and 40\%, respectively. This pattern suggests a lower degree of disruptiveness in their patented AI innovations. The evidence supports the interpretation that U.S. patents are, on average, of higher quality than their Chinese and South Korean counterparts. Indeed, despite having a smaller number of patents overall, U.S. patents receive more citations, include a higher proportion of international references, and display a faster and more pervasive citation rate. Interestingly, EU patents show an even faster citation rate than those from Japan and a similar residual share of uncited patents. This indicates that, although the quantitative contribution of EU-patented AI innovations is relatively modest, their quality or technological disruptiveness ranks among the highest globally.

\subsection{A gravity model approach}

To understand the determinants of international AI knowledge flows, we apply the gravity model framework originally developed for international trade analysis \citep{anderson_gravity_2003}. The gravity model posits that bilateral trade flows between two entities are proportional to their economic masses and inversely related to the distance between them. In our context, we adapt this framework to analyze how AI patent citations-a proxy for knowledge spillovers-flow between countries.
The theoretical foundation rests on the idea that knowledge, like goods, faces resistance when crossing borders. Geographic distance, cultural differences, and institutional barriers all create "friction" that impedes the flow of technological knowledge \citep{jaffe_geographic_1993, keller_geographic_2002}. Simultaneously, countries with larger AI innovation capacity (mass) generate and attract more knowledge flows, creating a gravitational pull analogous to Newton's law of universal gravitation. 

\noindent Our gravity model specification takes the following multiplicative form:

\begin{align}
\small
\mathbb{E}[citations_{ijt} \mid X_{ijt}] 
&= \exp\!\Big( \alpha 
+ \beta_{1} \ln(AI\;patents_{it}) 
+ \beta_{2} \ln(AI\;patents_{jt}) \notag \\
&\quad + \beta_{3} AI\;technological\;proximity_{ijt} 
+ \gamma X_{ijt} +  \mu_{t}\Big)
\label{eq: gravity}
\end{align}

\noindent where $citations_{ijt}$ represents is the number of AI patent citations from country $i$ to country $j$ in year $t$, $AI\;patents_{it}$ and $AI\;patents_{jt}$ represent the stock of AI patents granted to firms in origin and destination countries in year $t$, $AI\;technological\;proximity_{ijt}$ is the measure exposed in Section \ref{sec: tech_prox}, and $X_{ijt}$ is a vector of bilateral cultural and institutional variables, and $\mu_{t}$ are time fixed effects.\\

\noindent We sourced our bilateral measures from the CEPII gravity database \citep{conte_cepii_2022}. Specifically, for $distance_{ij}$ we used the population-weighted distance using harmonic mean\footnote{The population-weighted harmonic distance is theoretically preferred among other distance measures as it accounts for the spatial distribution of economic activity within countries and corresponds to the distance elasticity typically found in trade studies \citep{head_illusory_2010}.}.
Among $X_{ijt}$ we include $distance_{ij}$ measuring the spatial resistance to knowledge flows between countries, the dummy variables $common\_language_{ij}$, for countries sharing a common official language, $common\_legal_{ij}$, for countries sharing common legal origins after the post-socialist transition period, $colonial\_relations_{ij}$, signaling if countries are or were in colonial relationship after the year 1945, $contiguous_{ij}$, indicating contiguous countries, and $eu_{ij}$ indicating if both countries are part of the European Union. We also include $common\_religion_{ij}$, a religious proximity index, and the natural logarithm of each country GDP. In addition we leverage data on research and development (R\&D) expenditure  as percentage of the GDP from the World Bank. \\

\noindent Since according to \cite{silva_log_2006}, gravity models face an inherent econometric challenge where the heteroskedastic nature of residuals, exacerbated by taking logarithms of the dependent variable, violates key OLS assumptions and produces biased coefficient estimates. We implement their solution of employing Poisson pseudo-maximum likelihood estimation (PPML), which remains robust to this particular form of heteroskedasticity. Finally, to handle the presence of zeros we add 0.0001 to each variable before taking the logarithmic transformation. \\

\subsubsection{Results and discussion} 
\label{sec: gravity_res}
Table \ref{tab: gravity_res} reports our PPML estimates of bilateral AI patent citation flows for different sets of covariates. Across all model specifications, the results confirm the strong role of both geographic and economic factors in shaping international knowledge diffusion in artificial intelligence. In the baseline gravity framework (column 1), geographic distance exerts a consistently negative and highly significant effect on cross-border citations. A 1\% increase in distance between two countries reduces expected AI knowledge flows by roughly 0.6\%, confirming the existence of strong spatial frictions in the diffusion of technological knowledge. This coefficient remains remarkably stable across alternative model specifications, underscoring the robustness of the geographic distance effect. Cultural proximity variables show heterogeneous effects. Common official language emerges as a strong facilitator of AI knowledge diffusion, increasing expected citations by about 73\% ($e^{0.552}-1$) in the most parsimonious model. Legal frameworks similarity and former colonial ties does not exert a significant effect, while common religion becomes significant and negative only when controlling for additional country characteristics, indicating that cultural alignment matters mostly through shared language rather than institutional heritage. Contiguity generally displays negative associations, suggesting that geographic proximity alone is not sufficient to generate citation links once linguistic and technological factors are considered. \\

\noindent When controlling for economic size and R\&D investments (column 2) we notice how both $ln(GDP_i)$ and $ln(GDP_j)$ are positive and highly significant, with elasticities around to 1.4. This implies that larger economies both generate and attract more AI knowledge flows, consistent with the idea that economic mass fosters absorptive capacity and innovation linkages. Common religion turns negative and significant. As expected  R\&D investments at a country level show positive and significant coefficients, suggesting a central role of such financial accounts in both citing and cited countries, in fostering patented AI knowledge diffusion. \\

\begin{table}[htpb]
    \centering
        \caption{Gravity model results.}
        \label{tab: gravity_res}
\resizebox{.75\textwidth}{!}{%
\begin{tabular}{lcccc} \hline
 & (1) & (2) & (3) & (4) \\ \hline
 &  &  &  &  \\
$(log\;of)\;distance_{ijt}$ & -0.609*** & -0.731*** & -0.076* & -0.099** \\
 & (0.086) & (0.090) & (0.042) & (0.045) \\
$common\_language_{ij}$ & 0.552** & 1.033*** & 0.442** & 0.446** \\
 & (0.242) & (0.318) & (0.192) & (0.184) \\
$common\_legal_{ij}$ & -0.325 & -0.317 & -0.005 & -0.060 \\
 & (0.258) & (0.244) & (0.152) & (0.139) \\
$common\_religion_{ij}$ & 0.466 & -1.802** & 1.290** & 1.659*** \\
 & (0.919) & (0.768) & (0.506) & (0.559) \\
$colonial\_relations_{ij}$ & -0.235 & -0.115 & -0.369 & -0.376 \\
 & (0.219) & (0.275) & (0.251) & (0.260) \\
$contiguous_{ij}$ & -1.023*** & -1.010*** & -0.212 & -0.338* \\
 & (0.264) & (0.337) & (0.197) & (0.193) \\
$regional\;trade\;agreement_{ij}$ & 0.003 & 0.134 & -0.307*** & -0.290** \\
 & (0.181) & (0.238) & (0.114) & (0.135) \\
$(log\;of)\;GDP_{it}$ &  & 1.467*** & 0.128 & 0.121 \\
 &  & (0.089) & (0.083) & (0.080) \\
$(log\;of)\;GDP_{jt}$ &  & 1.372*** & 0.231*** & 0.236*** \\
 &  & (0.076) & (0.061) & (0.065) \\
$(log\;of)\;GDP\;per\;capita_{it}$ &  & -0.355** & 0.272** & 0.244** \\
 &  & (0.144) & (0.115) & (0.114) \\
$(log\;of)\;GDP\;per\;capita_{jt}$ &  & 0.082 & 0.619*** & 0.581*** \\
 &  & (0.145) & (0.097) & (0.093) \\
$R\&D\;investments_{i}$ &  & 0.571*** & -0.049 & -0.062 \\
 &  & (0.136) & (0.077) & (0.076) \\
$R\&D\;investments_{j}$ &  & 0.472*** & 0.088 & 0.077 \\
 &  & (0.128) & (0.056) & (0.058) \\
$(log\;of)\;AI\;patents_{it}$ &  &  & 0.788*** & 0.769*** \\
 &  &  & (0.046) & (0.045) \\
$(log\;of)\;AI\;patents_{jt}$ &  &  & 0.645*** & 0.614*** \\
 &  &  & (0.052) & (0.054) \\
$AI\;technological\;proximity_{ij}$ &  &  & 0.060*** & 0.058*** \\
 &  &  & (0.006) & (0.006) \\
$eu_{i}\;(origin)$ &  &  &  & -0.151 \\
 &  &  &  & (0.192) \\
$eu_{j}\;(destination)$ &  &  &  & -0.188 \\
 &  &  &  & (0.137) \\
$eu_{ij}$ &  &  &  & -0.363 \\
 &  &  &  & (0.357) \\
  &  &  &  &  \\
\textit{Controls:} & & & &\\
Time FE & \checkmark & \checkmark & \checkmark & \checkmark \\
Origin country FE & \checkmark &  - & - & - \\
Destination country FE & \checkmark & - & - & - \\
& & & & \\ \hline
Observations & 43,321 & 29,327 & 29,327 & 29,327 \\
 Pseudo R-square & 0.967 & 0.956 & 0.985 & 0.986 \\ \hline \hline
\end{tabular}
}
    \begin{tablenotes}
        \footnotesize 
        \item Note: The table reports Poisson pseudo-maximum likelihood estimation results of our gravity model applied to AI patents citations. Errors are clustered at the dyad level.  *** p$<$0.01, ** p$<$0.05, * p$<$0.1.
    \end{tablenotes}
\end{table}

\noindent Introducing AI specific variables (column 3), fundamentally alters the gravity structure of AI knowledge flows, absorbing much of the variation previously captured by language and distance/contiguity. Common religion turns positive and significant, with an effect of 263\% ($e^{1.29}-1$) -indicating that, conditional on similar technological trajectory, shared religious background facilitates cross-border AI knowledge diffusion, whereas in earlier specifications it likely captured less innovative regional clusters, yielding a spurious negative sign. The elasticity of AI patenting activity is substantial: a 1 percent increase in the number of AI patents in either country raises expected citation flows by about 0.7\%, confirming that AI-intensive systems are more likely to exchange knowledge internationally. The inclusion of these variables also attenuates the distance effect, suggesting that part of the apparent geographic friction reflects differences in innovative capacity. The coefficient for technological proximity is positive and significant, indicating that countries with similar AI technological portfolios exhibit stronger citation linkages-evidence of path-dependent and domain-specific spillovers. Specifically, a 1\% increase in technological proximity, translates in a modest 0.06\% higher expected citations. Regional trade agreements becomes significant but negative. The coefficients for GDP per capita change sign and become positive for both citing and cited countries. After controlling for AI patent stocks, the negative coefficient on per-capita GDP in the citing country turn positive ($\simeq0.3\%$) as the (larger) one for the cited one ($\simeq0.6\%$), suggesting that knowledge tends to flow from richer to poorer economies more easily, highlighting the asymmetric nature of AI knowledge diffusion. Finally, R\&D investment is no longer significant. Once the scale of AI patenting is included, R\&D expenditure adds little explanatory power for the source of citations-its effect is likely already embodied in patent output. Overall, these shifts confirm that AI knowledge diffusion is primarily driven by technological capability and specialization, rather than by geography, culture, or legacy institutional ties. \\ 

\noindent In the final specification (column 4), which introduces dummy variables for EU membership of both citing and cited countries, the coefficients on the EU variables don't show statistical significance. This suggests that, once differences in countries’ economic size, R\&D intensity, and AI patenting capacity are accounted for, belonging to the European Union does not exert an independent effect on the intensity of cross-border AI knowledge flows. At the same time, the coefficient on contiguity becomes negative and significant again, indicating that outside the EU framework, geographic adjacency often reflects institutional or linguistic barriers that hinder rather than facilitate AI knowledge diffusion. The stronger negative elasticity of distance further supports this interpretation: when regional integration and cultural proximity are held constant, physical distance re-emerges as the main friction in the international transmission of AI-related knowledge. \\

\noindent Taken together, the gravity estimations reveal a nuanced geography of global AI knowledge diffusion. While geographic distance remains the most persistent and robust barrier to international AI citation flows, its effect diminishes once differences in technological capacity are taken into account, underscoring the central role of innovation capabilities in shaping global knowledge linkages. Economic mass and AI patent intensity emerge as the dominant facilitators of cross-border knowledge exchange, highlighting that larger and more technologically specialized countries occupy pivotal positions in the international AI innovation network. Cultural and institutional factors exert more heterogeneous influences: shared language consistently promotes diffusion, whereas religion, colonial ties, and legal similarity matter only conditionally and lose explanatory power when technological proximity is accounted for. Finally, the absence of a significant EU effect once technological variables are included suggests that formal regional integration alone does not guarantee stronger AI knowledge flows-what ultimately drives international diffusion is technological specialization and absorptive capacity rather than geography or political affiliation. Overall, the findings point to an increasingly capability-driven global innovation landscape, where distance still matters, but technological dimension matters more.

\subsubsection{Robustness check}

A potential concern with the estimation of international AI knowledge flows through the gravity model is the presence of sample selection bias. While the PPML estimator efficiently handles heteroskedasticity and the presence of zero flows, it implicitly assumes that all observed zeros arise from random variation around a continuous latent process. However, many country pairs record no forward AI patent citations because no effective knowledge link exists between them, that is, because the fixed costs of establishing a citation channel exceed potential gains. In such cases, the process determining whether a citation link exists may differ systematically from the process determining its intensity, generating non-random sample selection \citep{heckman_sample_1979}. \\

\begin{table}[htpb]
    \centering
        \caption{Gravity model results with Heckman correction.}
        \label{tab: gravity_res_heckman}
\resizebox{.7\textwidth}{!}{%
\begin{tabular}{lcccc} \hline
 & (1) & (2) & (3) & (4) \\ \hline
 &  &  &  &  \\
$(log\;of)\;distance_{ijt}$ & -0.609*** & -0.711*** & -0.077* & -0.097** \\
 & (0.086) & (0.092) & (0.043) & (0.045) \\
$common\_language_{ij}$ & 0.560** & 0.913*** & 0.422** & 0.411** \\
 & (0.246) & (0.335) & (0.195) & (0.186) \\
$common\_legal_{ij}$ & -0.322 & -0.285 & -0.000 & -0.052 \\
 & (0.259) & (0.259) & (0.153) & (0.139) \\
$common\_religion_{ij}$ & 0.462 & -1.512* & 1.305** & 1.717*** \\
 & (0.919) & (0.799) & (0.514) & (0.575) \\
$colonial\_relations_{ij}$ & -0.235 & -0.115 & -0.368 & -0.381 \\
 & (0.217) & (0.281) & (0.250) & (0.258) \\
$contiguous_{ij}$ & -1.023*** & -1.069*** & -0.225 & -0.359* \\
 & (0.267) & (0.348) & (0.200) & (0.195) \\
$regional\;trade\;agreement_{ij}$ & -0.006 & 0.160 & -0.300*** & -0.285** \\
 & (0.184) & (0.241) & (0.114) & (0.134) \\
$(log\;of)\;GDP_{it}$ &  & 1.408*** & 0.132 & 0.123 \\
 &  & (0.094) & (0.084) & (0.080) \\
$(log\;of)\;GDP_{jt}$ &  & 1.321*** & 0.233*** & 0.238*** \\
 &  & (0.080) & (0.061) & (0.066) \\
$(log\;of)\;GDP\;per\;capita_{it}$ &  & -0.337** & 0.265** & 0.232** \\
 &  & (0.149) & (0.115) & (0.115) \\
$(log\;of)\;GDP\;per\;capita_{jt}$ &  & 0.081 & 0.613*** & 0.569*** \\
 &  & (0.146) & (0.097) & (0.093) \\
$R\&D\;investments_{i}$ &  & 0.506*** & -0.051 & -0.067 \\
 &  & (0.142) & (0.078) & (0.076) \\
$R\&D\;investments_{j}$ &  & 0.413*** & 0.085 & 0.070 \\
 &  & (0.134) & (0.057) & (0.058) \\
$(log\;of)\;AI\;patents_{it}$ &  &  & 0.779*** & 0.755*** \\
 &  &  & (0.046) & (0.045) \\
$(log\;of)\;AI\;patents_{jt}$ &  &  & 0.638*** & 0.601*** \\
 &  &  & (0.053) & (0.055) \\
$AI\;technological\;proximity_{ij}$ &  &  & 0.060*** & 0.058*** \\
 &  &  & (0.006) & (0.006) \\
$eu_{i}\;(origin)$ &  &  &  & -0.180 \\
 &  &  &  & (0.193) \\
$eu_{j}\;(destination)$ &  &  &  & -0.216 \\
 &  &  &  & (0.137) \\
$eu_{ij}$ &  &  &  & -0.342 \\
 &  &  &  & (0.359) \\
IMR & 0.938*** & -2.008*** & 0.156 & -0.056 \\
 & (0.317) & (0.628) & (0.202) & (0.220) \\
  &  &  &  &  \\
\textit{Controls:} & & & &\\
Time FE & \checkmark & \checkmark & \checkmark & \checkmark \\
Origin country FE & \checkmark &  - & - & - \\
Destination country FE & \checkmark & - & - & - \\
& & & & \\ \hline
Observations & 10,225 & 7,510 & 7,510 & 7,510 \\
 Pseudo R-square & 0.957 & 0.944 & 0.981 & 0.982 \\ \hline \hline
\end{tabular}
}
\begin{tablenotes}
    \footnotesize 
    \item Note: The table reports Poisson pseudo-maximum likelihood estimation results of our gravity model applied to AI patents citations, correcting for Heckman selection bias. Errors are clustered at the dyad level.  *** p$<$0.01, ** p$<$0.05, * p$<$0.1.
\end{tablenotes}
\end{table}

\noindent To assess the robustness of our main results, we follow the correction strategy proposed by \citet{helpman_estimating_2008}, who develop a structural model of international trade flows that extends Heckman selection framework to the context of gravity equations. Although the formation of knowledge flows through patent citations differs conceptually from the formation of trade flows between countries-since citations do not face the same institutional or cost barriers that constrain physical trade-we adopt their two-step correction as a diagnostic tool. This allows us to test whether unobserved factors jointly affect the probability that a cross-country citation link exists and the intensity of citation exchanges, even if the underlying selection mechanism may be weaker than in the case of trade. \\

\noindent In the first stage, we estimate a probit model capturing the likelihood that country $i$ cites patents from country $j$ in year $t$:
\begin{equation}
\Pr(y_{ijt} > 0) = \Phi(Z_{ijt}\gamma),
\label{eq:heckman1}
\end{equation}
where $y_{ijt}$ denotes the observed number of forward AI patent citations from $i$ to $j$, $\Phi(\cdot)$ is the cumulative distribution function of the standard normal, and $Z_{ijt}$ is a vector of determinants of citation link formation, including bilateral distance, common language, common religion, colonial history, contiguity, bilateral trade agreements, and time, origin and destination fixed effects. \\

\noindent From equation~(\ref{eq:heckman1}), we compute the inverse Mills ratio (IMR):
\begin{equation}
\lambda_{ijt} = \frac{\phi(Z_{ijt}\hat{\gamma})}{\Phi(Z_{ijt}\hat{\gamma})},
\label{eq:imr}
\end{equation}

\noindent where $\phi(\cdot)$ and $\Phi(\cdot)$ denote the standard normal density and cumulative distribution functions, respectively. The IMR represents the expected value of the unobserved error term conditional on selection into the positive-flow sample and serves as a correction term in the outcome equation. In the second stage, we then re-estimate the gravity equations on the positive-flow subsample using the PPML estimator augmented with the correction term:

\begin{equation}
E[citations_{ijt} \mid citations_{ijt} > 0, X_{ijt}] = 
\exp\!\left( \alpha + \beta' X_{ijt} + \delta \lambda_{ijt} + \mu_t \right)
\label{eq:heckman2}
\end{equation}

\noindent where $X_{ijt}$ includes the same set of gravity covariates and the coefficient $\delta$ captures the correlation between the unobserved determinants of citation link formation and citation intensity. A statistically significant $\delta$ would indicate the presence of selection bias in the uncorrected model. \\
 
\noindent This two-step approach conceptually decomposes international AI citation dynamics into two stages: (i) an extensive margin determining whether a citation link is established between countries, and (ii) an intensive margin determining the magnitude of citations once such a link exists. Following \citet{helpman_estimating_2008}, this decomposition allows us to test whether factors such as geographical distance, language similarity, or technological proximity influence both margins simultaneously. \\

\noindent The results of the Heckman-corrected estimation (Table \ref{tab: gravity_res_heckman}) confirm the robustness of the baseline PPML estimates. The coefficients remain highly consistent in both sign and magnitude, indicating that sample selection bias plays a limited role in shaping international AI citation flows. The inverse Mills ratio is significant for the first two specifications, suggesting that while non-random link formation exists, its economic relevance is modest once technological and economic factors are controlled for. Crucially, distance, AI patenting capacity, and technological proximity retain their explanatory power, reaffirming that the main drivers of AI knowledge diffusion operate along the intensive margin rather than being distorted by unobserved selection on the extensive one. This robustness check therefore strengthens confidence in the baseline interpretation: AI knowledge spillovers are primarily governed by distance, cultural affinity, economic size and technological capability specialization, rather than by selection effects or missing links in the citation network.


\section{Conclusions and final remarks}
\label{sec: conclusion}

This paper maps the global landscape of AI patenting between 2010 and 2023, providing a granular view of the geography, sectors, and organizational forms driving patented innovation,  with particular attention to Europe’s position within emerging international innovation dynamics. By integrating the Stanford AI Index dataset with firm- and ownership-level information from ORBIS IP, we have shown how AI patented innovation is simultaneously global in reach and highly concentrated in practice. China dominates in terms of patent counts, while the United States leads in terms of citation impact, highlighting a dual structure where quantity and quality are not aligned. Moreover, the asymmetry between national firms and multinationals underscores the dual role of organizational types: while national firms provide numerical breadth, multinational corporations concentrate technological frontier activity. \\

\noindent Our sectoral analysis reveals the centrality of ICT, manufacturing, and education, with universities and research institutes emerging as disproportionately influential contributors to AI patents stock. Measures of revealed comparative advantage and technological proximity further demonstrated that specialization in AI does not always coincide with absolute leadership: smaller countries such as Israel and Ireland show strong relative specialization despite lower absolute volumes. This suggests that global AI patented innovation is characterized not only by concentration but also by strategic diversification across different innovation systems. \\

\noindent The empirical evidence from the gravity model underscores that distance still constrains AI knowledge diffusion, yet its impact weakens significantly when accounting for differences in innovation capacity and specialization. Economic mass, R\&D intensity, and particularly AI patenting activity emerge as the dominant facilitators of cross-border citation flows, confirming that technologically advanced countries are pivotal nodes in the international knowledge network. \\

\noindent A key contribution of this work is its examination of Europe’s position in this evolving global system, and the findings are threefold. First, despite EU’s large number of AI-patenting entities, its collective patent output remains fragmented and comparatively limited in scale. The European Union’s aggregated RCA value - comparable to Japan’s but below the world average - suggests that AI is not yet a central focus of its patented innovation portfolio. Nonetheless, Europe retains distinct strengths. Countries such as Ireland, Sweden, and Finland demonstrate relatively high AI specialization and technological proximity to the U.S. frontier, pointing to the existence of competitive regional niches within the broader European ecosystem. These findings indicate that while Europe lags in aggregate patent production, it possesses the institutional and human capital foundations required to strengthen its AI innovation capacity - provided it can improve cross-border collaboration and accelerate the translation of research into industrial applications. Second, European AI patents exhibit relatively high technological quality - as reflected by their faster citation dynamics and lower share of uncited patents compared to Asian peers - suggesting that European innovation retains strong scientific and technological depth. Third, the lack of a significant EU effect in the gravity model reveals persistent fragmentation within the continent’s AI ecosystem: formal integration and shared regulatory frameworks have yet to produce the kind of cohesive innovation network observed in the United States or China.
For Europe, these findings raise critical questions about the effectiveness of ongoing initiatives, such as the Coordinated Plan on AI\footnote{digital-strategy.ec.europa.eu/en/library/coordinated-plan-artificial-intelligence} and the Chips Act\footnote{digital-strategy.ec.europa.eu/en/policies/european-chips-act}, in fostering continental technological sovereignty. From a policy perspective, this underscores the strategic importance of promoting such target within the EU, not merely through protectionist measures, but by deepening cross-border collaborations, harmonizing R\&D incentives, and consolidating Europe’s industrial base around AI-intensive technologies. In this regard, the asymmetry of AI patents production in the Education sector, with respect to global leaders, suggests the importance of fostering private and public investments into local Universities and research centers. Furthermore, strengthening technological proximity and absorptive capacity across member states would enable Europe to act as a more autonomous and globally competitive pole in the evolving AI innovation system. In sum, AI knowledge diffusion is increasingly driven by technological specialization and innovation capacity, not geography. Europe’s challenge lies not in bridging physical distance, but in transforming its fragmented innovation potential into a unified source of technological power. \\

\noindent Future research should explore the firm-level performance implications of AI patenting, specifically distinguishing the dynamics of emerging domains like generative AI and autonomous robotics. Given the variations in sectoral distribution, it is also crucial to investigate productivity dynamics across different fields, particularly in education and manufacturing. For the latter, distinguishing between the internal deployment versus external commercialization of AI-patented technologies will be essential to understanding the changing division of labor in this sector.

\section*{Data availability statement}
The data supporting the findings of this study are from PATSTAT Global (EPO) and ORBIS IP database (Bureau Van Dijk). Access to these data requires a valid license and they are not publicly available. Natural language processing code used in this study are available from the authors upon reasonable request.

\section*{Acknowledgements}
\textit{The authors acknowledge that while this work is based on their original ideas and analysis, the writing process was aided by AI tools. Specifically, Mistral's LLM was used to tighten and edit initial drafts.}

\pagebreak

\newpage

\setlength\bibsep{0.5pt}
\footnotesize
\bibliographystyle{elsarticle-harv}
\bibliography{bibliography.bib}

\newpage


\appendix
\section*{Appendix A: Tables and graphs}
\label{sec:appendix_tab_graph}
\setcounter{table}{0}
\renewcommand{\thetable}{A\arabic{table}}
\setcounter{figure}{0}
\renewcommand{\thefigure}{A\arabic{figure}}

This section contains additional material and the complete distributions of figures mentioned in each section of the paper.

\begin{table}[h]
    \centering
        \caption{Missing data for firm-level variables.}
        \label{tab: missing_orbis}
    \begin{tabular}{lc}
    \hline
        Variable & Missing share \\ \hline
        Date of incorporation & 11\% \\ 
        NACE & 10\% \\ 
        Parent company NACE & 14\% \\ 
        N. employees & 43\% \\ 
\hline \hline
\end{tabular}
\begin{tablenotes}
\footnotesize 
\item Note: The table reports percentage shares of missing information for firm-level variables on AI applicants, retrieved from ORBIS IP.
\end{tablenotes}
\end{table}

\begin{figure}[h]
    \begin{center}
    \caption{Density distribution of AI-patentees age}
    \label{fig: age_dist}
    \includegraphics[scale=.7]{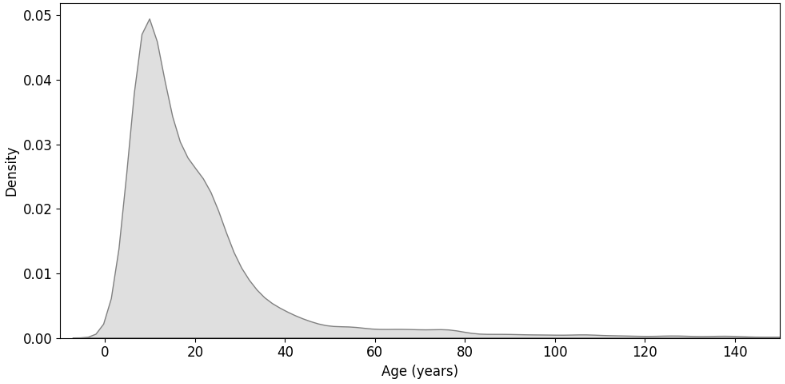}
    \end{center}
    \begin{tablenotes}
\footnotesize 
\item Note: The figure represent the density distribution of the age of patentees with at least one granted AI-patent. Period 2010-2023.
\end{tablenotes}
\end{figure}

\begin{table}[h]
    \centering
        \caption{Distribution of AI-firms and their patent over 2-digits NACE manufacturing sectors.}
        \label{tab: dist_firm_nace_manufacture}
\resizebox{1\textwidth}{!}{%
    \begin{tabular}{C{2cm} L{8cm} cccc} \hline
        NACE & Description & N. firms & \% of total (firms) & N. AI-patents & \% of total (AI-patents) \\ 
    \hline
        10 & Food products & 149 & 1.2 & 242 & 0.3 \\ 
        11 & Beverages & 19 & 0.1 & 41 & <0.1 \\ 
        12 & Tobacco products & 30 & 0.2 & 145 & 0.2 \\ 
        13 & Textiles & 82 & 0.6 & 115 & 0.1 \\ 
        14 & Wearing apparel & 54 & 0.4 & 81 & 0.1 \\ 
        15 & Leather and related products & 23 & 0.2 & 70 & 0.1 \\ 
        16 & Wood and of products of wood and cork & 38 & 0.3 & 61 & 0.1 \\ 
        17 & Paper and paper products & 70 & 0.5 & 141 & 0.1 \\ 
        18 & Printing and reproduction of recorded media & 128 & 1.0 & 381 & 0.4 \\ 
        19 & Coke and refined petroleum products & 47 & 0.4 & 417 & 0.4 \\ 
        20 & Chemicals & 319 & 2.5 & 1,451 & 1.5 \\ 
        21 & Basic pharmaceutical & 312 & 2.4 & 1,091 & 1.1 \\ 
        22 & Rubber and plastic products & 220 & 1.7 & 363 & 0.4 \\ 
        23 & Othernon metallic mineral products & 113 & 0.9 & 199 & 0.2 \\ 
        24 & Basic metals & 171 & 1.3 & 626 & 0.7 \\ 
        25 & Fabricated metal products & 521 & 4.0 & 1,308 & 1.4 \\ 
        26 & Computer and electronics & 4,634 & 36.0 & 51,216 & 53.9 \\ 
        27 & Electrical equipment & 1,330 & 10.3 & 5,322 & 5.6 \\ 
        28 & Machinery and equipment & 2,111 & 16.4 & 8,414 & 8.9 \\ 
        29 & Motorvehicles and trailers & 673 & 5.2 & 15,223 & 16.0 \\ 
        30 & Other transport equipment & 388 & 3.0 & 3,312 & 3.5 \\ 
        31 & Furniture & 77 & 0.6 & 120 & 0.1 \\ 
        32 & Other manufacturing & 1,283 & 10.0 & 4,359 & 4.6 \\ 
        33 & Repair of machinery and equipment & 88 & 0.7 & 283 & 0.3 \\ 
\hline \hline
\end{tabular}
}
\begin{tablenotes}
\footnotesize 
\item Note: The table reports the distribution of firms with at least one granted AI-patent, and their respective patents production, over 2-digits NACE manufacturing sub-sectors. Percentage shares refers to the current specific sub-sample. Period 2010-2023.
\end{tablenotes}
\end{table}

\begin{figure}[h]
    \begin{center}
    \caption{Density distribution of AI-firms RCA}
    \label{fig: rca_dist}
    \includegraphics[scale=.8]{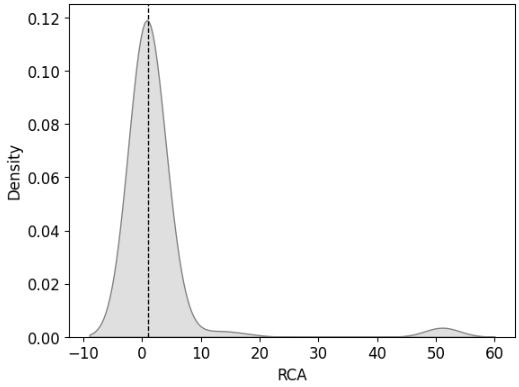}
    \end{center}
    \begin{tablenotes}
\footnotesize 
\item Note: The figure represent the density distribution of the RCA of firms with at least one granted AI-patent. Dashed line corresponds to value 1: threshold for relative advantage ($>$) or disadvantage ($<$). Period 2010-2023.
\end{tablenotes}
\end{figure}

\begin{figure}[h]
    \begin{center}
    \caption{N. of granted AI patents vs IPCs/CPCs involved - aggregation by country.}
    \label{fig: cpc_scatter}
    \includegraphics[scale=.6]{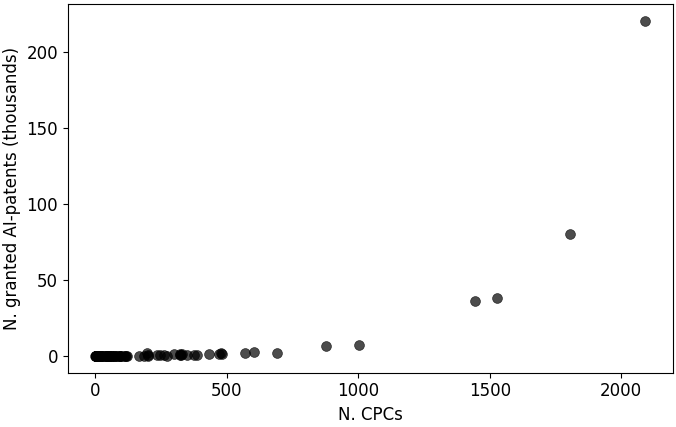}
    \end{center}
    \begin{tablenotes}
\footnotesize 
\item Note: The figure contains the scatter plot of the number of granted AI patents in a country vs. the number of 5-digits IPCs/CPCs domains related to such grants. Period 2010-2023.
\end{tablenotes}
\end{figure}

\begin{table}[h]
    \centering
        \caption{Descriptive statistics on forward citations distribution.}
        \label{tab: citation_stat}
    \begin{tabular}{lcc}
    \hline
        ~ & (1) & (2) \\ 
        ~ & Patent family & Country \\ \cline{2-3} 
        && \\
        Mean & 7 & 22,278 \\ 
        Median & 2 & 82 \\ 
        Std. dev. & 22 & 126,886 \\ 
        Max & 2,098 & 1,147,971 \\ 
        Min & 0 & 0 \\ 
\hline \hline
\end{tabular}
\begin{tablenotes}
\footnotesize 
\item Note: The table reports descriptive statistics about the distribution of the forward citations of AI patents among patent families (1) and countries (2).
\end{tablenotes}
\end{table}

\begin{figure}[h]
    \begin{center}
    \caption{Technological proximity heatmap.}
    \label{fig: proximity_heatmap}
    \includegraphics[scale=1.2]{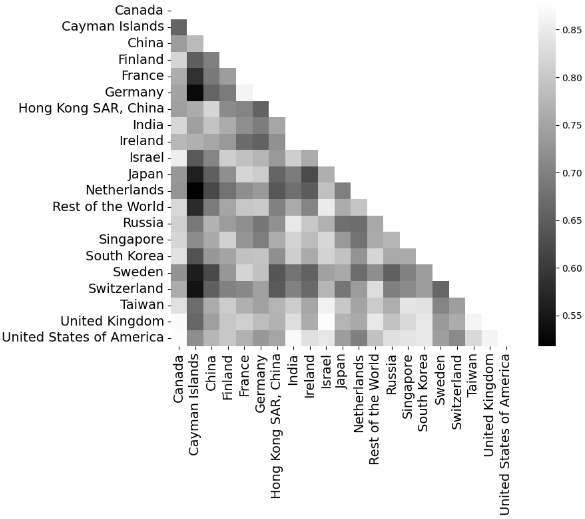}
    \end{center}
    \begin{tablenotes}
\footnotesize 
\item Note: The figure represent AI technological proximity heatmap by countries. In the category \textit{rest of the world} are aggregated countries which total AI patents production sum accounts for less than 1\% of the total. Period 2010-2023.
\end{tablenotes}
\end{figure}

\begin{table}[h]
    \centering
        \caption{Descriptive statistics of AI patents production by country - Part I}
        \label{tab: dist_firm_country_I}
\resizebox{1\textwidth}{!}{%
    \begin{tabular}{lcccccc}
    \hline
        Country & N. AI patents & Average per firm & N. AI patents S.D. & N. AI-firms & Average firm age & Firm age S.D. \\ \hline
        China & 220,463 & 9 & 77 & 24,664 & 17 & 12 \\ 
        United States of America & 80,371 & 8 & 111 & 9,961 & 31 & 37 \\ 
        South Korea & 38,054 & 6 & 59 & 6,237 & 16 & 12 \\ 
        Japan & 36,436 & 17 & 121 & 2,092 & 48 & 31 \\ 
        Germany & 6,971 & 8 & 53 & 825 & 40 & 39 \\ 
        Taiwan & 6,586 & 7 & 25 & 942 & 26 & 19 \\ 
        United Kingdom & 2,517 & 3 & 10 & 782 & 22 & 18 \\ 
        France & 2,143 & 6 & 20 & 340 & 31 & 21 \\ 
        Netherlands & 1,950 & 9 & 69 & 209 & 26 & 28 \\ 
        Canada & 1,771 & 3 & 9 & 540 & 39 & 45 \\ 
        Cayman Islands & 1,753 & 47 & 263 & 37 & 20 & 5 \\ 
        Israel & 1,609 & 3 & 8 & 560 & 16 & 11 \\ 
        Sweden & 1,439 & 7 & 25 & 197 & 29 & 27 \\ 
        India & 1,255 & 4 & 27 & 291 & 19 & 16 \\ 
        Ireland & 1,186 & 13 & 65 & 93 & 18 & 12 \\ 
        Switzerland & 1,163 & 5 & 10 & 241 & 33 & 33 \\ 
        Hong Kong SAR, China & 1,044 & 6 & 26 & 169 & 15 & 12 \\ 
        Finland & 936 & 7 & 34 & 139 & 26 & 25 \\ 
        Russia & 909 & 3 & 8 & 348 & 28 & 21 \\ 
        Singapore & 907 & 6 & 19 & 164 & 19 & 14 \\ 
        Australia & 686 & 2 & 3 & 372 & 21 & 22 \\ 
        Italy & 435 & 2 & 7 & 180 & 29 & 20 \\ 
        Denmark & 332 & 4 & 8 & 88 & 26 & 24 \\ 
        Spain & 321 & 2 & 3 & 156 & 26 & 20 \\ 
        Saudi Arabia & 317 & 20 & 40 & 16 & 36 & 18 \\ 
        Belgium & 311 & 3 & 5 & 115 & 28 & 25 \\ 
        Austria & 266 & 2 & 5 & 114 & 31 & 29 \\ 
        Barbados & 235 & 34 & 47 & 7 & 15 & 5 \\ 
        Malaysia & 206 & 5 & 11 & 38 & 24 & 12 \\ 
        Brazil & 160 & 2 & 3 & 77 & 33 & 17 \\ 
        Norway & 112 & 2 & 3 & 55 & 21 & 9 \\ 
        Czechia & 96 & 2 & 3 & 43 & 24 & 12 \\ 
        Cyprus & 95 & 5 & 10 & 20 & 15 & 11 \\ 
        Poland & 94 & 2 & 1 & 57 & 40 & 38 \\ 
        Luxembourg & 94 & 3 & 3 & 29 & 20 & 21 \\ 
        Portugal & 86 & 3 & 4 & 30 & 40 & 51 \\ 
        Puerto Rico & 68 & 6 & 13 & 12 & 24 & 43 \\ 
        New Zealand & 62 & 2 & 1 & 37 & 18 & 14 \\ 
        Turkiye & 60 & 2 & 3 & 25 & 29 & 16 \\ 
        South Africa & 51 & 2 & 2 & 29 & 16 & 12 \\ 
        Bulgaria & 42 & 1 & 1 & 34 & 20 & 14 \\ 
        Virgin Islands (British) & 37 & 3 & 3 & 13 & 32 & 0 \\ 
        Malta & 36 & 4 & 6 & 9 & 10 & 4 \\ 
        Chile & 30 & 2 & 3 & 12 & 55 & 71 \\ 
        Bermuda & 29 & 2 & 3 & 12 & 23 & 11 \\ 
        Lithuania & 29 & 2 & 2 & 13 & 17 & 10 \\ 
        Estonia & 28 & 3 & 3 & 9 & 14 & 6 \\ 
        United Arab Emirates & 27 & 2 & 1 & 15 & 21 & 14 \\ 
        Argentina & 27 & 3 & 6 & 8 & 60 & 10 \\ 
        Hungary & 26 & 1 & 1 & 19 & 30 & 41 \\ 
        Thailand & 25 & 2 & 3 & 11 & 31 & 15 \\ 
        Qatar & 22 & 6 & 6 & 4 & 26 & 18 \\ 
        Greece & 22 & 2 & 1 & 14 & 20 & 14 \\ 
        Ukraine & 22 & 3 & 3 & 8 & 23 & 6 \\ 
        Romania & 21 & 2 & 1 & 12 & 20 & 12 \\ 
\hline \hline
\end{tabular}
}
\begin{tablenotes}
\footnotesize 
\item Note: The table reports descriptive statistics about AI patents production and applicant age by country of the latter. Period 2010-2023.
\end{tablenotes}
\end{table}

\begin{table}[h]
    \centering
        \caption{Descriptive statistics of AI patents production by country - Part II}
        \label{tab: dist_firm_country_II}
\resizebox{1\textwidth}{!}{%
    \begin{tabular}{lcccccc}
    \hline
        Country & N. AI patents & Average per firm & N. AI patents S.D. & N. AI-firms & Average firm age & Firm age S.D. \\ \hline
        Croatia & 21 & 2 & 3 & 10 & 24 & 23 \\ 
        Slovenia & 21 & 1 & 1 & 16 & 33 & 16 \\ 
        Zimbabwe & 20 & 20 & 0 & 1 & 34 & 0 \\ 
        Mexico & 19 & 1 & 1 & 14 & 49 & 31 \\ 
        Slovakia & 18 & 2 & 2 & 8 & 52 & 29 \\ 
        Colombia & 15 & 1 & 1 & 12 & 36 & 32 \\ 
        Morocco & 12 & 2 & 3 & 5 & 42 & 47 \\ 
        Vietnam & 11 & 2 & 1 & 7 & 14 & 8 \\ 
        Ghana & 9 & 3 & 1 & 3 & 10 & 0 \\ 
        Liechtenstein & 8 & 3 & 2 & 3 & 65 & 39 \\ 
        Myanmar/Burma & 7 & 7 & 0 & 1 & 141 & 0 \\ 
        Liberia & 7 & 7 & 0 & 1 & 40 & 0 \\ 
        Philippines & 6 & 2 & 1 & 4 & 26 & 25 \\ 
        Kenya & 5 & 2 & 1 & 3 & 11 & 4 \\ 
        Jordan & 5 & 2 & 1 & 3 & 30 & 8 \\ 
        Serbia & 5 & 2 & 1 & 3 & 18 & 5 \\ 
        Iceland & 5 & 1 & 0 & 5 & 18 & 10 \\ 
        Paraguay & 5 & 2 & 1 & 3 & 17 & 10 \\ 
        Panama & 4 & 1 & 1 & 3 & 12 & 2 \\ 
        Oman & 4 & 2 & 1 & 2 & 20 & 2 \\ 
        United Republic of Tanzania & 4 & 4 & 0 & 1 & 17 & 0 \\ 
        Pakistan & 4 & 4 & 0 & 1 & 18 & 0 \\ 
        Sri Lanka & 3 & 1 & 0 & 3 & 0 & 0 \\ 
        Peru & 3 & 1 & 0 & 3 & 42 & 18 \\ 
        Indonesia & 3 & 1 & 0 & 3 & 53 & 0 \\ 
        Lesotho & 3 & 3 & 0 & 1 & 6 & 0 \\ 
        Belize & 3 & 2 & 1 & 2 & 6 & 0 \\ 
        Kuwait & 3 & 3 & 0 & 1 & 59 & 0 \\ 
        Macao SAR, China & 3 & 3 & 0 & 1 & 0 & 0 \\ 
        Nigeria & 3 & 2 & 1 & 2 & 34 & 9 \\ 
        Bosnia and Herzegovina & 3 & 3 & 0 & 1 & 27 & 0 \\ 
        Kazakhstan & 2 & 1 & 0 & 2 & 22 & 5 \\ 
        Zambia & 2 & 1 & 0 & 2 & 21 & 8 \\ 
        Latvia & 2 & 1 & 0 & 2 & 8 & 1 \\ 
        Papua New Guinea & 2 & 2 & 0 & 1 & 15 & 0 \\ 
        Ecuador & 2 & 2 & 0 & 1 & 0 & 0 \\ 
        Republic of Moldova & 2 & 1 & 0 & 2 & 24 & 9 \\ 
        Bahamas & 2 & 1 & 0 & 2 & 0 & 0 \\ 
        Belarus & 1 & 1 & 0 & 1 & 0 & 0 \\ 
        Trinidad and Tobago & 1 & 1 & 0 & 1 & 0 & 0 \\ 
        Uganda & 1 & 1 & 0 & 1 & 0 & 0 \\ 
        Bolivia & 1 & 1 & 0 & 1 & 12 & 0 \\ 
        Supranational & 1 & 1 & 0 & 1 & 0 & 0 \\ 
        Angola & 1 & 1 & 0 & 1 & 0 & 0 \\ 
        Algeria & 1 & 1 & 0 & 1 & 9 & 0 \\ 
        Gibraltar & 1 & 1 & 0 & 1 & 6 & 0 \\ 
        Samoa & 1 & 1 & 0 & 1 & 0 & 0 \\ 
        Seychelles & 1 & 1 & 0 & 1 & 0 & 0 \\ 
        Reunion (France) & 1 & 1 & 0 & 1 & 16 & 0 \\ 
        Monaco & 1 & 1 & 0 & 1 & 18 & 0 \\ 
        Malawi & 1 & 1 & 0 & 1 & 23 & 0 \\ 
        Greenland (Denmark) & 1 & 1 & 0 & 1 & 15 & 0 \\ 
        Mauritius & 1 & 1 & 0 & 1 & 13 & 0 \\ 
        Lebanon & 1 & 1 & 0 & 1 & 159 & 0 \\ 
        Costa Rica & 1 & 1 & 0 & 1 & 21 & 0 \\ 
        Antigua and Barbuda & 1 & 1 & 0 & 1 & 0 & 0 \\ 
\hline \hline
\end{tabular}
}
\begin{tablenotes}
\footnotesize 
\item Note: The table reports descriptive statistics about AI patents production and applicant age by country of the latter. Period 2010-2023.
\end{tablenotes}
\end{table}

\begin{table}[h]
    \centering
        \caption{Descriptive statistics by parent company country - Part I}
        \label{tab: dist_firm_country_guo_I}
\resizebox{1\textwidth}{!}{%
    \begin{tabular}{lcccccc}
    \hline
Country parent & N. AI patents & Average per firm & N. AI patents S.D. & N. AI-firms & Average firm age & Firm age S.D. \\ \hline
        China & 207,678 & 9 & 72 & 24,057 & 17 & 12 \\ 
        United States of America & 76,676 & 8 & 112 & 9,824 & 29 & 35 \\ 
        Japan & 38,976 & 16 & 114 & 2,370 & 48 & 32 \\ 
        South Korea & 38,285 & 6 & 59 & 6,226 & 16 & 12 \\ 
        Cayman Islands & 11,862 & 34 & 264 & 352 & 16 & 10 \\ 
        Germany & 7,530 & 9 & 53 & 842 & 45 & 46 \\ 
        Taiwan & 6,841 & 7 & 24 & 1,005 & 25 & 19 \\ 
        Hong Kong & 2,867 & 17 & 59 & 173 & 17 & 14 \\ 
        Netherlands & 2,533 & 11 & 66 & 239 & 31 & 37 \\ 
        United Kingdom & 2,241 & 3 & 9 & 713 & 24 & 27 \\ 
        France & 1,983 & 5 & 19 & 370 & 35 & 32 \\ 
        Virgin Islands (British) & 1,746 & 16 & 49 & 111 & 15 & 8 \\ 
        Canada & 1,672 & 3 & 10 & 501 & 36 & 42 \\ 
        Ireland & 1,401 & 11 & 58 & 124 & 34 & 34 \\ 
        Israel & 1,295 & 3 & 6 & 510 & 16 & 12 \\ 
        India & 1,287 & 5 & 28 & 281 & 18 & 16 \\ 
        Sweden & 1,232 & 5 & 21 & 236 & 28 & 27 \\ 
        Switzerland & 1,098 & 4 & 10 & 265 & 38 & 36 \\ 
        Finland & 988 & 8 & 36 & 130 & 25 & 24 \\ 
        Russia & 714 & 2 & 3 & 338 & 28 & 22 \\ 
        Australia & 605 & 2 & 3 & 341 & 20 & 22 \\ 
        Italy & 369 & 2 & 5 & 166 & 28 & 22 \\ 
        Hong Kong SAR, China & 348 & 4 & 9 & 93 & 12 & 8 \\ 
        Singapore & 346 & 3 & 5 & 123 & 20 & 14 \\ 
        Saudi Arabia & 324 & 15 & 36 & 21 & 36 & 22 \\ 
        Spain & 301 & 2 & 4 & 142 & 26 & 20 \\ 
        Denmark & 301 & 4 & 8 & 76 & 29 & 27 \\ 
        Bermuda & 280 & 5 & 11 & 54 & 23 & 12 \\ 
        Austria & 280 & 2 & 5 & 114 & 31 & 30 \\ 
        Belgium & 239 & 2 & 3 & 104 & 24 & 18 \\ 
        Luxembourg & 228 & 5 & 13 & 50 & 26 & 19 \\ 
        Curaçao & 152 & 10 & 21 & 15 & 55 & 30 \\ 
        Brazil & 128 & 2 & 3 & 66 & 34 & 19 \\ 
        Malaysia & 125 & 4 & 7 & 33 & 25 & 12 \\ 
        Norway & 121 & 2 & 4 & 56 & 24 & 21 \\ 
        Czechia & 93 & 3 & 3 & 37 & 26 & 13 \\ 
        Barbados & 92 & 23 & 25 & 4 & 13 & 3 \\ 
        Poland & 83 & 2 & 1 & 51 & 42 & 40 \\ 
        New Zealand & 77 & 2 & 3 & 38 & 19 & 14 \\ 
        United Arab Emirates & 68 & 3 & 5 & 23 & 21 & 12 \\ 
        South Africa & 61 & 2 & 2 & 33 & 18 & 13 \\ 
        Puerto Rico & 51 & 2 & 1 & 26 & 35 & 53 \\ 
        Cyprus & 47 & 2 & 3 & 20 & 13 & 10 \\ 
        Turkey & 43 & 2 & 3 & 18 & 32 & 19 \\ 
        Bulgaria & 42 & 1 & 1 & 34 & 20 & 14 \\ 
        Thailand & 28 & 3 & 3 & 11 & 28 & 14 \\ 
        Argentina & 27 & 3 & 6 & 8 & 60 & 10 \\ 
        Estonia & 24 & 3 & 3 & 7 & 16 & 5 \\ 
        Hungary & 22 & 1 & 1 & 16 & 33 & 44 \\ 
        Qatar & 22 & 6 & 6 & 4 & 26 & 18 \\ 
        Ukraine & 21 & 3 & 3 & 7 & 25 & 4 \\ 
        Slovenia & 21 & 1 & 1 & 16 & 33 & 16 \\ 
        Zimbabwe & 20 & 20 & 0 & 1 & 34 & 0 \\ 
        Croatia & 20 & 2 & 3 & 9 & 27 & 23 \\ 
        Mexico & 20 & 1 & 1 & 15 & 48 & 30 \\ 
        Greece & 20 & 2 & 1 & 13 & 18 & 11 \\ 
\hline \hline
\end{tabular}
}
\begin{tablenotes}
\footnotesize 
\item Note: The table reports descriptive statistics about AI patents production and applicant age by country of the parent company. Period 2010-2023.
\end{tablenotes}
\end{table}

\begin{table}[h]
    \centering
        \caption{Descriptive statistics by parent company country - Part II}
        \label{tab: dist_firm_country_guo_II}
\resizebox{1\textwidth}{!}{%
    \begin{tabular}{lcccccc}
    \hline
Country parent & N. AI patents & Average per firm & N. AI patents S.D. & N. AI-firms & Average firm age & Firm age S.D. \\ \hline
        Lithuania & 19 & 2 & 1 & 10 & 19 & 10 \\ 
        Portugal & 19 & 3 & 4 & 6 & 39 & 46 \\ 
        Chile & 18 & 2 & 2 & 11 & 55 & 71 \\ 
        Slovakia & 17 & 2 & 2 & 7 & 56 & 29 \\ 
        Romania & 16 & 2 & 1 & 8 & 17 & 13 \\ 
        Colombia & 15 & 1 & 1 & 12 & 36 & 32 \\ 
        Vietnam & 13 & 2 & 1 & 8 & 16 & 7 \\ 
        Morocco & 12 & 2 & 3 & 5 & 42 & 47 \\ 
        Liechtenstein & 10 & 2 & 1 & 6 & 44 & 37 \\ 
        Mauritius & 10 & 2 & 2 & 5 & 24 & 8 \\ 
        Turkiye & 9 & 2 & 2 & 5 & 14 & 4 \\ 
        Ghana & 9 & 3 & 1 & 3 & 10 & 0 \\ 
        Malta & 9 & 1 & 0 & 8 & 15 & 9 \\ 
        Myanmar/Burma & 7 & 7 & 0 & 1 & 141 & 0 \\ 
        Philippines & 7 & 1 & 1 & 5 & 24 & 22 \\ 
        Liberia & 7 & 7 & 0 & 1 & 40 & 0 \\ 
        Panama & 5 & 1 & 0 & 4 & 12 & 2 \\ 
        Jordan & 5 & 2 & 1 & 3 & 30 & 8 \\ 
        Kenya & 5 & 2 & 1 & 3 & 11 & 4 \\ 
        Paraguay & 5 & 2 & 1 & 3 & 17 & 10 \\ 
        Bahamas & 4 & 1 & 1 & 3 & 13 & 0 \\ 
        Seychelles & 4 & 1 & 1 & 3 & 14 & 7 \\ 
        Oman & 4 & 2 & 1 & 2 & 20 & 2 \\ 
        Pakistan & 4 & 4 & 0 & 1 & 18 & 0 \\ 
        United Republic of Tanzania & 4 & 4 & 0 & 1 & 17 & 0 \\ 
        Nigeria & 3 & 2 & 1 & 2 & 34 & 9 \\ 
        Bosnia and Herzegovina & 3 & 3 & 0 & 1 & 27 & 0 \\ 
        Gibraltar & 3 & 1 & 0 & 3 & 23 & 15 \\ 
        Iceland & 3 & 1 & 0 & 3 & 14 & 8 \\ 
        Sri Lanka & 3 & 1 & 0 & 3 & 0 & 0 \\ 
        Kuwait & 3 & 3 & 0 & 1 & 59 & 0 \\ 
        Macao SAR, China & 3 & 3 & 0 & 1 & 0 & 0 \\ 
        Lesotho & 3 & 3 & 0 & 1 & 6 & 0 \\ 
        Belize & 3 & 2 & 1 & 2 & 6 & 0 \\ 
        Papua New Guinea & 2 & 2 & 0 & 1 & 15 & 0 \\ 
        Peru & 2 & 1 & 0 & 2 & 48 & 22 \\ 
        Latvia & 2 & 1 & 0 & 2 & 8 & 1 \\ 
        Kazakhstan & 2 & 1 & 0 & 2 & 22 & 5 \\ 
        Indonesia & 2 & 1 & 0 & 2 & 24 & 0 \\ 
        Ecuador & 2 & 2 & 0 & 1 & 0 & 0 \\ 
        Zambia & 1 & 1 & 0 & 1 & 15 & 0 \\ 
        Lebanon & 1 & 1 & 0 & 1 & 159 & 0 \\ 
        Serbia & 1 & 1 & 0 & 1 & 22 & 0 \\ 
        Samoa & 1 & 1 & 0 & 1 & 0 & 0 \\ 
        Republic of Moldova & 1 & 1 & 0 & 1 & 31 & 0 \\ 
        Moldova & 1 & 1 & 0 & 1 & 18 & 0 \\ 
        Monaco & 1 & 1 & 0 & 1 & 18 & 0 \\ 
        Malawi & 1 & 1 & 0 & 1 & 23 & 0 \\ 
        Algeria & 1 & 1 & 0 & 1 & 9 & 0 \\ 
        Angola & 1 & 1 & 0 & 1 & 0 & 0 \\ 
        Anguilla & 1 & 1 & 0 & 1 & 30 & 0 \\ 
        Antigua and Barbuda & 1 & 1 & 0 & 1 & 0 & 0 \\ 
        Supranational & 1 & 1 & 0 & 1 & 0 & 0 \\ 
        Belarus & 1 & 1 & 0 & 1 & 0 & 0 \\ 
        Bolivia & 1 & 1 & 0 & 1 & 12 & 0 \\ 
        Trinidad and Tobago & 1 & 1 & 0 & 1 & 0 & 0 \\ 
        Uganda & 1 & 1 & 0 & 1 & 0 & 0 \\ 
\hline \hline
\end{tabular}
}
\begin{tablenotes}
\footnotesize 
\item Note: The table reports descriptive statistics about AI patents production and applicant age by country of the parent company. Period 2010-2023.
\end{tablenotes}
\end{table}

\begin{table}[h]
    \centering
        \caption{Distribution of AI-firms over CPC classes - Part I.}
        \label{tab: dist_firm_cpc_1}
\resizebox{.9\textwidth}{!}{%
    \begin{tabular}{C{1cm} L{17cm} C{2.5cm} C{1.8cm}}
    \hline
        Code & Description & N. AI patents & \% of total \\ \hline
        G06 & Computing; calculating; counting & 345,676 & 43.6 \\ 
        H04 & Electric communication technique & 73,970 & 9.3 \\ 
        Y02 & Technologies or applications for mitigation or adaptation against climate change & 56,238 & 7.1 \\ 
        G01 & Measuring; testing & 45,957 & 5.8 \\ 
        G10 & Musical instruments; acoustics & 43,328 & 5.5 \\ 
        G16 & Information and communication technology [ict] specially adapted for specific application fields & 29,580 & 3.7 \\ 
        A61 & Medical or veterinary science; hygiene & 25,966 & 3.3 \\ 
        G05 & Controlling; regulating & 23,959 & 3.0 \\ 
        G08 & Signalling & 21,637 & 2.7 \\ 
        B60 & Vehicles in general & 19,186 & 2.4 \\ 
        G07 & Checking-devices & 8,479 & 1.1 \\ 
        G09 & Educating; cryptography; display; advertising; seals & 8,139 & 1.0 \\ 
        B25 & Hand tools; portable power-driven tools; tool handles & 8,047 & 1.0 \\ 
        H02 & Generation, conversion, or distribution of electric power & 5,955 & 0.8 \\ 
        H01 & Electric elements & 5,160 & 0.7 \\ 
        Y04 & Information or communication technologies having an impact on other technology areas & 4,886 & 0.6 \\ 
        A63 & Sports; games; amusements & 4,175 & 0.5 \\ 
        G02 & Optics & 4,015 & 0.5 \\ 
        A01 & Agriculture;forestry; animal husbandry; hunting;trapping; fishing & 3,307 & 0.4 \\ 
        G11 & Information storage & 3,228 & 0.4 \\ 
        F16 & Engineering elements or units; general measures for effecting and maintaining effective functioning of machines & 3,098 & 0.4 \\ 
        H05 & Electric techniques not otherwise provided for & 2,566 & 0.3 \\ 
        B65 & Conveying; packing; storing; handling thin or filamentary material & 2,487 & 0.3 \\ 
        B64 & Aircraft; aviation; cosmonautics & 2,396 & 0.3 \\ 
        G03 & Photography; cinematography; analogous techniques using waves other than optical waves & 2,334 & 0.3 \\ 
        B62 & Land vehicles for travelling otherwise than on rails & 2,243 & 0.3 \\ 
        C12 & Biochemistry; beer; spirits; wine; vinegar; microbiology; enzymology; mutation or genetic engineering & 2,227 & 0.3 \\ 
        A47 & Furniture;domestic articles or appliances; coffee mills; spice mills;suction cleaners in general & 1,883 & 0.2 \\ 
        H03 & Electronic circuitry & 1,879 & 0.2 \\ 
        B07 & Separating solids from solids; sorting & 1,720 & 0.2 \\ 
        B23 & Machine tools; metal-working not otherwise provided for & 1,687 & 0.2 \\ 
        F24 & Heating; ranges; ventilating & 1,579 & 0.2 \\ 
        E21 & Earth drilling; mining & 1,535 & 0.2 \\ 
        H10 & Semiconductor devices; electric solid-state devices not otherwise provided for & 1,278 & 0.2 \\ 
        Y10 & Technical subjects covered by former uspc & 1,092 & 0.1 \\ 
        B66 & Hoisting; lifting; hauling & 1,086 & 0.1 \\ 
        B08 & Cleaning & 1,063 & 0.1 \\ 
        F21 & Lighting & 959 & 0.1 \\ 
        B01 & Physical or chemical processes or apparatus in general & 902 & 0.1 \\ 
        E05 & Locks; keys; window or door fittings; safes & 745 & 0.1 \\ 
        B29 & Working of plastics; working of substances in a plastic state in general & 700 & 0.1 \\ 
        F02 & Internal-combustion engines; hot-gas or combustion-product engine plants & 676 & 0.1 \\ 
        E01 & Construction of roads, railways, or bridges & 670 & 0.1 \\ 
        B61 & Railways & 659 & 0.1 \\ 
        F03 & Machines or engines for liquids; wind, spring, weight, or miscellaneous motors & 642 & 0.1 \\ 
        B41 & Printing; lining machines; typewriters; stamps & 612 & 0.1 \\ 
        B63 & Ships or other waterborne vessels; related equipment & 579 & 0.1 \\ 
        F05 & Indexing schemes relating to engines or pumps in various subclasses of classes f01-f04 & 560 & 0.1 \\ 
        E02 & Hydraulic engineering; foundations; soil shifting & 487 & 0.1 \\ 
        F04 & Engines or pumps & 483 & 0.1 \\ 
        E04 & Building & 467 & 0.1 \\ 
        B33 & Additive manufacturing technology & 450 & 0.1 \\ 
        F25 & Refrigeration or cooling; combined heating and refrigeration systems; heat pump systems & 428 & 0.1 \\ 
        B05 & Spraying or atomising in general; applying liquids or other fluent materials to surfaces, in general & 408 & 0.1 \\ 
        C02 & Treatment of water, waste water, sewage, or sludge & 401 & 0.1 \\ 
        A45 & Hand or travelling articles & 390 & 0.0 \\ 
        B21 & Mechanical metal-working without essentially removing material; punching metal & 385 & 0.0 \\ 
        A62 & Life-saving; fire-fighting & 356 & 0.0 \\ 
        D06 & Treatment of textiles or the like; laundering; flexible materials not otherwise provided for & 351 & 0.0 \\ 
        F41 & Weapons & 338 & 0.0 \\ 
        F17 & Storing or distributing gases or liquids & 323 & 0.0 \\ 
        F01 & Machines or engines in general; engine plants in general; steam engines & 308 & 0.0 \\ 
        A23 & Foods or foodstuffs; treatment thereof, not covered by other classes & 300 & 0.0 \\ 
        B22 & Casting; powder metallurgy & 290 & 0.0 \\ 
        B24 & Grinding; polishing & 269 & 0.0 \\ 
\hline \hline
\end{tabular}
}
\begin{tablenotes}
\footnotesize 
\item Note: The table reports the distribution of firms with at least one granted AI-patent, over 3-digits CPC classes. Period 2010-2023.
\end{tablenotes}
\end{table}

\begin{table}[h]
    \centering
        \caption{Distribution of AI-firms over CPC classes - Part II.}
        \label{tab: dist_firm_cpc_2}
\resizebox{.9\textwidth}{!}{%
    \begin{tabular}{C{1cm} L{17cm} C{2.5cm} C{1.8cm}}
    \hline
        Code & Description & N. AI patents & \% of total \\ \hline
        B82 & Nanotechnology & 246 & 0.0 \\ 
        C07 & Organic chemistry & 241 & 0.0 \\ 
        G04 & Horology & 226 & 0.0 \\ 
        E06 & Doors, windows, shutters, or roller blinds, in general; ladders & 198 & 0.0 \\ 
        A41 & Wearing apparel & 197 & 0.0 \\ 
        E03 & Water supply; sewerage & 193 & 0.0 \\ 
        C21 & Metallurgy of iron & 188 & 0.0 \\ 
        G21 & Nuclear physics; nuclear engineering & 173 & 0.0 \\ 
        B42 & Bookbinding; albums; files; special printed matter & 163 & 0.0 \\ 
        F23 & Combustion apparatus; combustion processes & 158 & 0.0 \\ 
        C40 & Combinatorial chemistry; libraries, e.g. chemical libraries, dna libraries & 153 & 0.0 \\ 
        B26 & Hand cutting tools; cutting; severing & 142 & 0.0 \\ 
        C09 & Dyes; paints, polishes, natural resins, adhesives; compositions not otherwise provided for & 135 & 0.0 \\ 
        A43 & Footwear & 135 & 0.0 \\ 
        A24 & Tobacco; cigars; cigarettes; smokers' requisites & 128 & 0.0 \\ 
        C23 & Coating metallic material; coating material with metallic material; chemical surface treatment & 125 & 0.0 \\ 
        B02 & Crushing, pulverising, or disintegrating; preparatory treatment of grain for milling & 118 & 0.0 \\ 
        B28 & Working cement, clay, or stone & 116 & 0.0 \\ 
        C25 & Electrolytic or electrophoretic processes; apparatus therefor & 115 & 0.0 \\ 
        A42 & Headwear & 111 & 0.0 \\ 
        C08 & Organic macromolecular compounds; their preparation or chemical working-up & 108 & 0.0 \\ 
        B67 & Opening or closing bottles, jars or similar containers; liquid handling & 99 & 0.0 \\ 
        F26 & Drying & 98 & 0.0 \\ 
        B32 & Layered products & 97 & 0.0 \\ 
        A44 & Haberdashery; jewellery & 97 & 0.0 \\ 
        F27 & Furnaces; kilns; ovens; retorts & 95 & 0.0 \\ 
        F28 & Heat exchange in general & 94 & 0.0 \\ 
        C01 & Inorganic chemistry & 93 & 0.0 \\ 
        C22 & Metallurgy; ferrous or non-ferrous alloys; treatment of alloys or non-ferrous metals & 89 & 0.0 \\ 
        B43 & Writing or drawing implements; bureau accessories & 87 & 0.0 \\ 
        C10 & Petroleum, gas or coke industries, technical gases containing carbon monoxide; fuels; lubricant; peat & 81 & 0.0 \\ 
        F42 & Ammunition; blasting & 76 & 0.0 \\ 
        B09 & Disposal of solid waste; reclamation of contaminated soil & 73 & 0.0 \\ 
        F15 & Fluid-pressure actuators; hydraulics or pneumatics in general & 73 & 0.0 \\ 
        D01 & Natural or man-made threads or fibres; spinning & 68 & 0.0 \\ 
        B03 & Separation of solid materials using liquids or using pneumatic tables or jigs & 65 & 0.0 \\ 
        A22 & Butchering; meat treatment; processing poultry or fish & 63 & 0.0 \\ 
        F22 & Steam generation & 63 & 0.0 \\ 
        C04 & Cements; concrete; artificial stone; ceramics; refractories & 57 & 0.0 \\ 
        C03 & Glass; mineral or slag wool & 55 & 0.0 \\ 
        B27 & Working or preserving wood or similar material; nailing or stapling machines in general & 53 & 0.0 \\ 
        D05 & Sewing; embroidering; tufting & 45 & 0.0 \\ 
        B06 & Generating or transmitting mechanical vibrations in general & 45 & 0.0 \\ 
        A46 & Brushware & 44 & 0.0 \\ 
        B30 & Presses & 43 & 0.0 \\ 
        B81 & Microstructural devices or systems, e.g. micromechanical devices & 40 & 0.0 \\ 
        D04 & Braiding; lace-making; knitting; trimmings; non-woven fabrics & 40 & 0.0 \\ 
        B44 & Decorative arts & 39 & 0.0 \\ 
        D03 & Weaving & 36 & 0.0 \\ 
        D21 & Paper-making; production of cellulose & 36 & 0.0 \\ 
        A21 & Baking; edible doughs & 33 & 0.0 \\ 
        C30 & Crystal growth & 32 & 0.0 \\ 
        B31 & Making paper articles; working paper & 31 & 0.0 \\ 
        D10 & Indexing scheme associated with sublasses of section d, relating to textiles & 19 & 0.0 \\ 
        C05 & Fertilisers; manufacture thereof & 17 & 0.0 \\ 
        D02 & Yarns; mechanical finishing of yarns or ropes; warping or beaming & 16 & 0.0 \\ 
        C11 & Animal or vegetable oils, fats, fatty substances or waxes; fatty acids therefrom; detergents; candles & 13 & 0.0 \\ 
        G12 & Instrument details & 11 & 0.0 \\ 
        B04 & Centrifugal apparatus or machines for carrying-out physical or chemical processes & 10 & 0.0 \\ 
        C14 & Skins; hides; pelts; leather & 8 & 0.0 \\ 
        A99 & Subject matter not otherwise provided for in section a & 7 & 0.0 \\ 
        C06 & Explosive; matches & 5 & 0.0 \\ 
        C13 & Sugar industry & 3 & 0.0 \\ 
        B68 & Saddlery; upholstery & 2 & 0.0 \\ 
        D07 & Ropes; cables other than electric & 1 & 0.0 \\ 
        G99 & Subject matter not otherwise provided for in section g & 1 & 0.0 \\ 
\hline \hline
\end{tabular}
}
\begin{tablenotes}
\footnotesize 
\item Note: The table reports the distribution of firms with at least one granted AI-patent, over 3-digits CPC classes. Period 2010-2023.
\end{tablenotes}
\end{table}

\begin{table}[h]
    \centering
        \caption{Countries with relative comparative \underline{advantage} in AI patents production.}
        \label{tab: rca_high}
\resizebox{.7\textwidth}{!}{%
    \begin{tabular}{lcccc}
    \hline
        Country & N.AI patents & N.Patents & \% of total & RCA \\ \hline
        United States of America & 80,371 & 933,528 & 9 & 5.15 \\ 
        Ireland & 1,186 & 13,311 & 9 & 4.52 \\ 
        Cayman Islands & 1,753 & 20,419 & 9 & 4.36 \\ 
        Qatar & 22 & 261 & 8 & 4.27 \\ 
        India & 1,255 & 15,256 & 8 & 4.18 \\ 
        Israel & 1,609 & 20,579 & 8 & 3.97 \\ 
        Barbados & 235 & 3,090 & 8 & 3.85 \\ 
        Puerto Rico & 68 & 944 & 7 & 3.65 \\ 
        Liberia & 7 & 111 & 6 & 3.19 \\ 
        Slovenia & 21 & 351 & 6 & 3.03 \\  
        Cyprus & 95 & 1,857 & 5 & 2.59 \\ 
        Vietnam & 11 & 215 & 5 & 2.59 \\ 
        Estonia & 28 & 590 & 5 & 2.4 \\ 
        Myanmar/Burma & 7 & 161 & 4 & 2.2 \\ 
        Singapore & 907 & 20,936 & 4 & 2.2 \\ 
        Virgin Islands (British) & 37 & 877 & 4 & 2.14 \\ 
        South Korea & 38,054 & 965,745 & 4 & 2.09 \\ 
        Malaysia & 206 & 4,993 & 4 & 2.09 \\ 
        Malta & 36 & 884 & 4 & 2.06 \\ 
        Saudi Arabia & 317 & 8,050 & 4 & 2.0 \\ 
        Canada & 1,771 & 49,634 & 4 & 1.81 \\ 
        Finland & 936 & 29,367 & 3 & 1.62 \\ 
        Morocco & 12 & 395 & 3 & 1.54 \\ 
        Ghana & 9 & 299 & 3 & 1.52 \\ 
        Thailand & 25 & 839 & 3 & 1.51 \\ 
        Turkiye & 60 & 2,023 & 3 & 1.5 \\ 
        Chile & 30 & 1,014 & 3 & 1.5 \\ 
        United Kingdom & 2,517 & 86,009 & 3 & 1.49 \\ 
        Portugal & 86 & 2,940 & 3 & 1.48 \\ 
        Jordan & 5 & 174 & 3 & 1.46 \\ 
        Lithuania & 29 & 1,045 & 3 & 1.41 \\ 
        Sweden & 1,439 & 53,020 & 3 & 1.38 \\ 
        Taiwan & 6,586 & 265,484 & 2 & 1.26 \\ 
        Netherlands & 1,950 & 79,111 & 2 & 1.25 \\ 
        Croatia & 21 & 868 & 2 & 1.23 \\ 
        Ukraine & 22 & 927 & 2 & 1.2 \\ 
        Greece & 22 & 977 & 2 & 1.14 \\ 
        Hong Kong SAR, China & 1,044 & 50,336 & 2 & 1.05 \\ 
\hline \hline
\end{tabular}
}
\begin{tablenotes}
\footnotesize 
\item Note: The table reports the list of countries with a relative comparative advantage in the production of AI-patented new technologies ($RCA > 1$). Period 2010-2023.
\end{tablenotes}
\end{table}

\begin{table}[h]
    \centering
        \caption{Countries with relative comparative \underline{disadvantage} in AI patents production.}
        \label{tab: rca_low}
\resizebox{.6\textwidth}{!}{%
    \begin{tabular}{lcccc}
    \hline
        Country & N.AI patents & N.Patents & \% of total & RCA \\ \hline
        Zimbabwe & 20 & 1,020 & 2 & 0.99 \\ 
        Costa Rica & 1 & 52 & 2 & 0.97 \\ 
        Slovakia & 18 & 991 & 2 & 0.92 \\ 
        United Arab Emirates & 27 & 1,508 & 2 & 0.91 \\ 
        Panama & 4 & 232 & 2 & 0.87 \\ 
        Denmark & 332 & 19,488 & 2 & 0.86 \\ 
        Japan & 36,436 & 2,112,676 & 2 & 0.86 \\ 
        Kuwait & 3 & 179 & 2 & 0.85 \\ 
        Switzerland & 1,163 & 73,594 & 2 & 0.8 \\ 
        Serbia & 5 & 322 & 2 & 0.79 \\ 
        France & 2,143 & 138,108 & 2 & 0.78 \\ 
        Germany & 6,971 & 459,585 & 2 & 0.76 \\ 
        Bulgaria & 42 & 2,946 & 1 & 0.72 \\ 
        Papua New Guinea & 2 & 142 & 1 & 0.71 \\ 
        Luxembourg & 94 & 6,758 & 1 & 0.7 \\ 
        Australia & 686 & 50,392 & 1 & 0.69 \\ 
        Belgium & 311 & 23,233 & 1 & 0.68 \\ 
        Spain & 321 & 24,308 & 1 & 0.67 \\ 
        Antigua and Barbuda & 1 & 78 & 1 & 0.65 \\ 
        Latvia & 2 & 158 & 1 & 0.64 \\ 
        Argentina & 27 & 2,169 & 1 & 0.63 \\ 
        Seychelles & 1 & 85 & 1 & 0.6 \\ 
        Hungary & 26 & 2,222 & 1 & 0.59 \\ 
        Norway & 112 & 9,637 & 1 & 0.59 \\ 
        New Zealand & 62 & 5,441 & 1 & 0.58 \\ 
        Belize & 3 & 267 & 1 & 0.57 \\ 
        South Africa & 51 & 4,733 & 1 & 0.55 \\ 
        Iceland & 5 & 478 & 1 & 0.53 \\ 
        Austria & 266 & 25,538 & 1 & 0.53 \\ 
        China & 220,463 & 14,653,708 & 2 & 0.47 \\ 
        Bermuda & 29 & 3,327 & 1 & 0.44 \\ 
        Gibraltar & 1 & 115 & 1 & 0.44 \\ 
        Pakistan & 4 & 471 & 1 & 0.43 \\ 
        Italy & 435 & 52,357 & 1 & 0.42 \\ 
        Monaco & 1 & 129 & 1 & 0.39 \\ 
        Colombia & 15 & 1,936 & 1 & 0.39 \\ 
        Mexico & 19 & 2,521 & 1 & 0.38 \\ 
        Macao SAR, China & 3 & 402 & 1 & 0.38 \\ 
        Philippines & 6 & 821 & 1 & 0.37 \\ 
        Brazil & 160 & 22,387 & 1 & 0.36 \\ 
        Nigeria & 3 & 459 & 1 & 0.33 \\ 
        Romania & 21 & 3,218 & 1 & 0.33 \\ 
        Russia & 909 & 161,465 & 1 & 0.28 \\ 
        Indonesia & 3 & 559 & 1 & 0.27 \\ 
        Czechia & 96 & 17,981 & 1 & 0.27 \\ 
        Mauritius & 1 & 209 & $<1$ & 0.24 \\ 
        Kazakhstan & 2 & 450 & $<1$ & 0.23 \\ 
        Kenya & 5 & 1,154 & $<1$ & 0.22 \\ 
        Poland & 94 & 31,386 & $<1$ & 0.15 \\ 
        Uganda & 1 & 404 & $<1$ & 0.13 \\ 
        Liechtenstein & 8 & 3,233 & $<1$ & 0.13 \\ 
        Bahamas & 2 & 1,150 & $<1$ & 0.09 \\ 
        Republic of Moldova & 2 & 1,438 & $<1$ & 0.07 \\ 
        Belarus & 1 & 790 & $<1$ & 0.06 \\ 
\hline \hline
\end{tabular}
}
\begin{tablenotes}
\footnotesize 
\item Note: The table reports the list of countries with a relative comparative advantage in the production of AI-patented new technologies ($RCA \leq 1$). Period 2010-2023.
\end{tablenotes}
\end{table}

\begin{table}[h]
    \centering
        \caption{Distribution of family level forward citations by country - Part I.}
        \label{tab: citations_dist_p1}
\resizebox{.8\textwidth}{!}{%
    \begin{tabular}{lcccccc}
    \hline
        Country & Tot. citations & Mean & Std. & Min & Max & \% of total \\ \hline
        United States of America & 1,147,971 & 18 & 44 & 1 & 2,098 & 46.8 \\ 
        China & 651,202 & 5 & 7 & 1 & 272 & 26.6 \\ 
        Japan & 212,548 & 9 & 15 & 1 & 450 & 8.7 \\ 
        South Korea & 125,618 & 6 & 12 & 1 & 470 & 5.1 \\ 
        Germany & 48,699 & 10 & 20 & 1 & 299 & 2 \\ 
        Canada & 26,886 & 19 & 40 & 1 & 568 & 1.1 \\ 
        United Kingdom & 24,905 & 13 & 26 & 1 & 292 & 1 \\ 
        Taiwan & 24,742 & 7 & 13 & 1 & 219 & 1 \\ 
        Israel & 24,474 & 19 & 33 & 1 & 422 & 1 \\ 
        Ireland & 18,975 & 18 & 48 & 1 & 969 & 0.8 \\ 
        Netherlands & 17,971 & 11 & 21 & 1 & 416 & 0.7 \\ 
        Switzerland & 16,294 & 20 & 49 & 1 & 394 & 0.7 \\ 
        France & 12,839 & 9 & 29 & 1 & 606 & 0.5 \\ 
        Sweden & 11,290 & 12 & 22 & 1 & 333 & 0.5 \\ 
        Cayman Islands & 9,940 & 6 & 10 & 1 & 186 & 0.4 \\ 
        Australia & 7,682 & 16 & 34 & 1 & 334 & 0.3 \\ 
        Hong Kong SAR, China & 7,480 & 8 & 12 & 1 & 120 & 0.3 \\ 
        India & 7,039 & 8 & 14 & 1 & 227 & 0.3 \\ 
        Puerto Rico & 6,861 & 108 & 136 & 1 & 394 & 0.3 \\ 
        Finland & 6,838 & 12 & 20 & 1 & 229 & 0.3 \\ 
        Singapore & 6,102 & 10 & 21 & 1 & 210 & 0.2 \\ 
        Barbados & 3,643 & 19 & 53 & 1 & 390 & 0.1 \\ 
        Italy & 3,394 & 13 & 27 & 1 & 283 & 0.1 \\ 
        Belgium & 3,174 & 17 & 63 & 1 & 780 & 0.1 \\ 
        Spain & 2,991 & 19 & 48 & 1 & 340 & 0.1 \\ 
        Russia & 2,634 & 5 & 10 & 1 & 215 & 0.1 \\ 
        Denmark & 2,450 & 11 & 18 & 1 & 170 & 0.1 \\ 
        Saudi Arabia & 2,255 & 9 & 15 & 1 & 136 & 0.1 \\ 
        Cyprus & 1,812 & 28 & 57 & 1 & 298 & 0.1 \\ 
        Brazil & 1,198 & 15 & 38 & 1 & 300 & 0 \\ 
        Austria & 1,177 & 8 & 11 & 1 & 70 & 0 \\ 
        Norway & 1,086 & 12 & 29 & 1 & 181 & 0 \\ 
        Malaysia & 918 & 9 & 12 & 1 & 94 & 0 \\ 
        South Africa & 837 & 25 & 43 & 1 & 190 & 0 \\ 
        New Zealand & 673 & 16 & 25 & 1 & 132 & 0 \\ 
        Portugal & 608 & 14 & 25 & 1 & 98 & 0 \\ 
        Bermuda & 526 & 21 & 34 & 1 & 129 & 0 \\ 
        Luxembourg & 467 & 7 & 11 & 1 & 48 & 0 \\ 
        Virgin Islands (British) & 452 & 14 & 15 & 1 & 76 & 0 \\ 
        Hungary & 333 & 17 & 18 & 1 & 72 & 0 \\ 
        Poland & 332 & 8 & 14 & 1 & 86 & 0 \\ 
        Turkiye & 262 & 7 & 8 & 1 & 41 & 0 \\ 
        Czechia & 232 & 6 & 9 & 1 & 54 & 0 \\ 
        Argentina & 219 & 9 & 13 & 1 & 60 & 0 \\ 
        Chile & 211 & 8 & 8 & 1 & 38 & 0 \\ 
        Romania & 202 & 18 & 21 & 1 & 66 & 0 \\ 
        Qatar & 189 & 11 & 8 & 1 & 25 & 0 \\ 
        Estonia & 188 & 8 & 14 & 1 & 72 & 0 \\ 
        Colombia & 183 & 15 & 21 & 1 & 76 & 0 \\
\hline \hline
\end{tabular}
}
\begin{tablenotes}
\footnotesize 
\item Note: The table reports the distribution of family level forward citations of AI patents, by country. Period 2010-2023.
\end{tablenotes}
\end{table}

\begin{table}[h]
    \centering
        \caption{Distribution of family level forward citations by country - Part II.}
        \label{tab: citations_dist_p2}
\resizebox{.75\textwidth}{!}{%
    \begin{tabular}{lcccccc}
    \hline
        Country & Tot. citations & Mean & Std. & Min & Max & \% of total \\ \hline 
        Thailand & 142 & 8 & 11 & 1 & 37 & 0 \\ 
        United Arab Emirates & 133 & 8 & 7 & 1 & 27 & 0 \\ 
        Mexico & 133 & 12 & 16 & 1 & 47 & 0 \\ 
        Greece & 124 & 9 & 11 & 1 & 40 & 0 \\ 
        Malta & 122 & 4 & 3 & 1 & 15 & 0 \\ 
        Iceland & 96 & 24 & 12 & 6 & 33 & 0 \\ 
        Zimbabwe & 68 & 7 & 7 & 1 & 23 & 0 \\ 
        Jordan & 51 & 17 & 6 & 13 & 24 & 0 \\ 
        Bulgaria & 48 & 5 & 6 & 1 & 21 & 0 \\ 
        Croatia & 47 & 3 & 4 & 1 & 16 & 0 \\ 
        Angola & 45 & 45 & 0 & 45 & 45 & 0 \\ 
        United Republic of Tanzania & 44 & 11 & 8 & 4 & 23 & 0 \\ 
        Philippines & 41 & 20 & 26 & 2 & 39 & 0 \\ 
        Oman & 39 & 9 & 9 & 2 & 20 & 0 \\ 
        Slovenia & 37 & 4 & 2 & 1 & 9 & 0 \\ 
        Kuwait & 35 & 17 & 20 & 3 & 32 & 0 \\ 
        Ukraine & 31 & 10 & 9 & 1 & 19 & 0 \\ 
        Ghana & 31 & 6 & 6 & 1 & 16 & 0 \\ 
        Lithuania & 25 & 1 & 1 & 1 & 4 & 0 \\ 
        Liberia & 24 & 4 & 2 & 1 & 9 & 0 \\ 
        Vietnam & 23 & 5 & 3 & 3 & 11 & 0 \\ 
        Algeria & 23 & 23 & 0 & 23 & 23 & 0 \\ 
        Indonesia & 21 & 10 & 2 & 9 & 12 & 0 \\ 
        Sri Lanka & 19 & 6 & 4 & 2 & 10 & 0 \\ 
        Greenland (Denmark) & 15 & 15 & 0 & 15 & 15 & 0 \\ 
        Kenya & 14 & 7 & 5 & 3 & 11 & 0 \\ 
        Myanmar/Burma & 14 & 2 & 1 & 1 & 4 & 0 \\ 
        Trinidad and Tobago & 14 & 14 & 0 & 14 & 14 & 0 \\ 
        Turkey & 12 & 12 & 0 & 12 & 12 & 0 \\ 
        Zambia & 10 & 5 & 5 & 1 & 9 & 0 \\ 
        Liechtenstein & 8 & 2 & 2 & 1 & 5 & 0 \\ 
        Bahamas & 7 & 3 & 3 & 1 & 6 & 0 \\ 
        Panama & 7 & 3 & 3 & 1 & 6 & 0 \\ 
        Belize & 7 & 2 & 1 & 1 & 4 & 0 \\ 
        Pakistan & 7 & 2 & 1 & 1 & 4 & 0 \\ 
        Republic of Moldova & 5 & 2 & 2 & 1 & 4 & 0 \\ 
        Antigua and Barbuda & 4 & 4 & 0 & 4 & 4 & 0 \\ 
        Slovakia & 4 & 2 & 0 & 2 & 2 & 0 \\ 
        Bosnia and Herzegovina & 3 & 1 & 0 & 1 & 1 & 0 \\ 
        Paraguay & 3 & 1 & 0 & 1 & 2 & 0 \\ 
        Gibraltar & 3 & 3 & 0 & 3 & 3 & 0 \\ 
        Nigeria & 3 & 1 & 0 & 1 & 1 & 0 \\ 
        Reunion (France) & 2 & 2 & 0 & 2 & 2 & 0 \\ 
        Kazakhstan & 2 & 2 & 0 & 2 & 2 & 0 \\ 
        Macao SAR, China & 2 & 1 & 0 & 1 & 1 & 0 \\ 
        Monaco & 1 & 1 & 0 & 1 & 1 & 0 \\ 
        Malawi & 1 & 1 & 0 & 1 & 1 & 0 \\ 
        Morocco & 1 & 1 & 0 & 1 & 1 & 0 \\ 
        Papua New Guinea & 1 & 1 & 0 & 1 & 1 & 0 \\ 
        Lebanon & 1 & 1 & 0 & 1 & 1 & 0 \\ 
        Costa Rica & 1 & 1 & 0 & 1 & 1 & 0 \\ 
        Seychelles & 1 & 1 & 0 & 1 & 1 & 0 \\ 
        Serbia & 1 & 1 & 0 & 1 & 1 & 0 \\ 
\hline \hline
\end{tabular}
}
\begin{tablenotes}
\footnotesize 
\item Note: The table reports the distribution of family level forward citations of AI patents, by country. Period 2010-2023.
\end{tablenotes}
\end{table}

\end{document}